\pdfoutput=1
\documentclass[apj,twocolumn,twocolappendix,numberedappendix]{openjournal}

\usepackage{amsmath}
\usepackage{booktabs}
\usepackage{multirow}
\usepackage{color}
\usepackage{soul}
\usepackage{threeparttable}
\usepackage{float}
\usepackage{graphicx}
\usepackage{CJK}
\usepackage{xspace}
\usepackage{afterpage}
\usepackage{placeins}
\usepackage{natbib}
\usepackage[breaklinks,colorlinks,citecolor=blue,urlcolor=blue,linkcolor=blue,filecolor=blue]{hyperref}

\usepackage[export]{adjustbox}
\usepackage{orcidlink}

\def\farcs{%
 \mbox{%
  \kern  0.13ex.%
  \kern -0.95ex\arcsec%
  \kern -0.1ex%
 }%
}%

\newcommand{\oiii}{[O\,{\sc iii}]}

\newcommand{\hst}{{\it HST}}

\newcommand{\jwst}{{\it JWST}}

\newcommand{\kms}{km\,s$^{-1}$\xspace}
\newcommand{\snh}{SN Helios\xspace}
\newcommand{\cluster}{RXC\,J0018.5+1626\xspace}
\newcommand{\zsn}{$z=3.34$\xspace}
\newcommand{\ha}{H$\alpha$\xspace}
\newcommand{\hb}{H$\beta$\xspace}
\newcommand{\hei}{He\,{\sc i}\,$\lambda$10830\xspace}

\defcitealias{Galbany2016}{[1]}
\defcitealias{Pastorello2009}{[2]}
\defcitealias{Faran2014}{[3]}
\defcitealias{Brown2014}{[4]}
\defcitealias{Dessart2008}{[5]}
\defcitealias{TsvetkovPavlyuk2014}{[6]}
\defcitealias{Yuan2016}{[7]}
\defcitealias{Huang2015}{[8]}
\defcitealias{Dhungana2016}{[9]}
\defcitealias{Leonard2002}{[10]}
\defcitealias{Hamuy2001}{[11]}
\defcitealias{Silverman2012}{[12]}
\defcitealias{Gutierrez2017}{[13]}
\defcitealias{YaronGalYam2012}{[14]}
\defcitealias{Valenti2014}{[15]}
\defcitealias{Childress2016}{[16]}

\shorttitle{A Multiply Imaged Type II Supernova at $z=3.34$}
\shortauthors{Fujimoto et al.}

\begin{document}

\title{
SN Helios: A Multiply Imaged Type II Supernova \\ 
Opening Time-Delay Cosmography beyond redshift of 3 
\vspace{-15mm}}

\author{
Seiji Fujimoto$^{1,2}$\footnotemark[*]\orcidlink{0000-0001-7201-5066},
Conor~Larison$^{3}$\orcidlink{0000-0003-2037-4619},
Justin D. R. Pierel$^{4}$\orcidlink{0000-0002-2361-7201},
Lukas J. Furtak$^{5,6}$\orcidlink{0000-0001-6278-032X},
Masamune Oguri$^{7,8}$\orcidlink{0000-0003-3484-399X},
Adi Zitrin$^{9}$\orcidlink{0000-0002-0350-4488},
Jose M. Diego$^{10}$\orcidlink{0000-0001-9065-3926},
Keren Sharon$^{11}$\orcidlink{0000-0002-7559-0864},
David~A.~Coulter$^{12,3}$\orcidlink{0000-0003-4263-2228},
Gabriel Brammer$^{13,14}$\orcidlink{0000-0003-2680-005X},
Vasily Kokorev$^{6}$\orcidlink{0000-0002-5588-9156},
Chelsea Nash$^{11}$\orcidlink{0009-0007-8190-7729},
Pedram Abedi$^{11}$\orcidlink{0009-0004-9243-3459},
Christa DeCoursey$^{15}$\orcidlink{0000-0002-4781-9078},
Armin Rest$^{4}$\orcidlink{0000-0002-4410-5387},
Andreas L. Faisst$^{16}$\orcidlink{0000-0002-9382-9832},
Brenda L. Frye$^{17}$\orcidlink{0000-0003-1625-8009},
Tiger Yu-Yang Hsiao$^{6,5}$\orcidlink{0000-0003-4512-8705},
Kohei Inayoshi$^{18}$\orcidlink{0000-0001-9840-4959},
Kotaro Kohno$^{19,20}$\orcidlink{0000-0002-4052-2394},
Jorryt Matthee$^{21}$\orcidlink{0000-0003-2871-127X},
Matteo Messa$^{22}$\orcidlink{0000-0003-1427-2456},
Yoshiaki Ono$^{23}$\orcidlink{0000-0001-9011-7605},
Fengwu Sun$^{24}$\orcidlink{0000-0002-4622-6617},
Eros Vanzella$^{22}$\orcidlink{0000-0002-5057-135X},
Rogier A. Windhorst$^{25}$\orcidlink{0000-0001-8156-6281},
Abdurro'uf$^{26}$\orcidlink{0000-0002-5258-8761},
Aadya Agrawal$^{27}$\orcidlink{0009-0008-1965-9012},
Joseph F. V. Allingham$^{28}$\orcidlink{0000-0003-2718-8640},
Yoshihisa Asada$^{29,30}$\orcidlink{0000-0003-3983-5438},
Hakim Atek$^{31}$\orcidlink{0000-0002-7570-0824},
Maru\v{s}a Brada\v{c}$^{32}$\orcidlink{0000-0001-5984-0395},
Larry D. Bradley$^{4}$\orcidlink{0000-0002-7908-9284},
Mateusz Bronikowski$^{33}$\orcidlink{0000-0002-1537-6911},
Dan Coe$^{34,35,36}$\orcidlink{0000-0001-7410-7669},
Christopher J. Conselice$^{37}$\orcidlink{0000-0003-1949-7638},
Pratika Dayal$^{38,39,40}$\orcidlink{0000-0001-8460-1564},
Eiichi Egami$^{41}$\orcidlink{0000-0003-1344-9475},
Michael Engesser$^{4}$\orcidlink{0000-0003-0209-674X},
Gavin Scott Farley$^{42}$\orcidlink{0009-0007-2578-9238},
Qinyue Fei$^{1}$\orcidlink{0000-0001-7232-5355},
Ori D. Fox$^{4}$\orcidlink{0000-0003-2238-1572},
Yoshinobu Fudamoto$^{7}$\orcidlink{0000-0001-7440-8832},
Miriam Golubchik$^{28}$\orcidlink{0000-0001-9411-3484},
Yuichi Harikane$^{23}$\orcidlink{0000-0002-6047-430X},
Gourav Khullar$^{43,44}$\orcidlink{0000-0002-3475-7648},
Tomokazu Kiyota$^{45,46}$\orcidlink{0009-0004-4332-9225},
Paulo A. A. Lopes$^{47}$\orcidlink{0000-0003-2540-7424},
Ray A. Lucas$^{4}$\orcidlink{0000-0003-1581-7825},
Guillaume Mahler$^{48}$\orcidlink{0000-0003-3266-2001},
Ashish Kumar Meena$^{49}$\orcidlink{0000-0002-7876-4321},
Rohan P. Naidu$^{50}$\orcidlink{0000-0003-3729-1684},
Gautham Narayan$^{27}$\orcidlink{0000-0001-6022-0484},
Ga\"el Noirot$^{4}$,
Pascal~A.~Oesch$^{51,52}$\orcidlink{0000-0001-5851-6649},
Masami Ouchi$^{53,23,54,55}$\orcidlink{0000-0002-1049-6658},
Jos\'e Mar\'ia Palencia$^{56}$\orcidlink{0000-0003-0942-817X},
Massimo Pascale$^{57}$\orcidlink{0000-0002-2282-8795},
Luke Robbins$^{58}$\orcidlink{0000-0002-6265-2675},
M.~R.~Siebert$^{4}$\orcidlink{0000-0003-2445-3891},
Tommaso Treu$^{57}$\orcidlink{0000-0002-8460-0390},
Roberta Tripodi$^{59}$\orcidlink{0000-0002-9909-3491},
Francesco Valentino$^{13,60}$\orcidlink{0000-0001-6477-4011},
Darach Watson$^{13,61}$\orcidlink{0000-0002-4465-8264},
Hayley Williams$^{25}$\orcidlink{0000-0002-1681-0767},
Hiroto Yanagisawa$^{23,62}$\orcidlink{0009-0006-6763-4245},
Anita Zanella$^{22}$\orcidlink{0000-0001-8600-7008}
}

\affiliation{$^{1}$David A. Dunlap Department of Astronomy and Astrophysics, University of Toronto, 50 St. George Street, Toronto, Ontario, M5S 3H4, Canada}
\affiliation{$^{2}$Dunlap Institute for Astronomy and Astrophysics, 50 St. George Street, Toronto, Ontario, M5S 3H4, Canada}
\affiliation{$^{3}$Space Telescope Science Institute, Baltimore, MD 21218, USA}
\affiliation{$^{4}$Space Telescope Science Institute, 3700 San Martin Drive, Baltimore, MD 21218, USA}
\affiliation{$^{5}$Cosmic Frontier Center, The University of Texas at Austin, Austin, TX 78712, USA}
\affiliation{$^{6}$Department of Astronomy, The University of Texas at Austin, Austin, TX 78712, USA}
\affiliation{$^{7}$Center for Frontier Science, Chiba University, 1-33 Yayoi-cho, Inage-ku, Chiba 263-8522, Japan}
\affiliation{$^{8}$Department of Physics, Graduate School of Science, Chiba University, 1-33 Yayoi-Cho, Inage-Ku, Chiba 263-8522, Japan}
\affiliation{$^{9}$Department of Physics, Ben-Gurion University of the Negev, P.O. Box 653, Beer-Sheva 8410501, Israel}
\affiliation{$^{10}$Instituto de F\'isica de Cantabria (CSIC-UC), Avda. Los Castros s/n. 39005, Santander, Spain}
\affiliation{$^{11}$Department of Astronomy, University of Michigan, 1085 South University Avenue, Ann Arbor, MI 48109, USA}
\affiliation{$^{12}$Physics and Astronomy Department, Johns Hopkins University, Baltimore, MD 21218, USA}
\affiliation{$^{13}$Cosmic Dawn Center (DAWN), Denmark}
\affiliation{$^{14}$Niels Bohr Institute, University of Copenhagen, Jagtvej 128, DK-2200 Copenhagen N, Denmark}
\affiliation{$^{15}$Steward Observatory, University of Arizona, 933 N. Cherry Ave, Tucson, AZ 85721, USA}
\affiliation{$^{16}$IPAC, California Institute of Technology, 1200 E. California Blvd., Pasadena, CA 91125, USA}
\affiliation{$^{17}$Department of Astronomy/Steward Observatory, University of Arizona, 933 N. Cherry Avenue, Tucson, AZ 85721, USA}
\affiliation{$^{18}$Kavli Institute for Astronomy and Astrophysics, Peking University, Beijing 100871, China}
\affiliation{$^{19}$Institute of Astronomy, Graduate School of Science, The University of Tokyo, 2-21-1 Osawa, Mitaka, Tokyo, 181-0015 Japan}
\affiliation{$^{20}$Research Center for the Early Universe, Graduate School of Science, The University of Tokyo, 7-3-1 Hongo, Bunkyo-ku, Tokyo 113-0033, Japan}
\affiliation{$^{21}$Institute of Science and Technology Austria (ISTA), Am Campus 1, 3400 Klosterneuburg, Austria}
\affiliation{$^{22}$INAF -- OAS, Osservatorio di Astrofisica e Scienza dello Spazio di Bologna, via Gobetti 93/3, I-40129 Bologna, Italy}
\affiliation{$^{23}$Institute for Cosmic Ray Research, The University of Tokyo, 5-1-5 Kashiwanoha, Kashiwa, Chiba 277-8582, Japan}
\affiliation{$^{24}$Center for Astrophysics $|$ Harvard \& Smithsonian, 60 Garden St., Cambridge, MA 02138, USA}
\affiliation{$^{25}$School of Earth and Space Exploration, Arizona State University, Tempe, AZ 85287-6004, USA}
\affiliation{$^{26}$Department of Astronomy, Indiana University, 727 East Third Street, Bloomington, IN 47405, USA}
\affiliation{$^{27}$Department of Astronomy, University of Illinois Urbana-Champaign, 1002 West Green Street, Urbana, IL 61801, USA}
\affiliation{$^{28}$Department of Physics, Ben-Gurion University of the Negev, P.O. Box 653, Be'er-Sheva 84105, Israel}
\affiliation{$^{29}$Dunlap Institute for Astronomy and Astrophysics, 50 St. George Street, Toronto, Ontario M5S 3H4, Canada}
\affiliation{$^{30}$Dunlap Fellow}
\affiliation{$^{31}$Institut d'Astrophysique de Paris, CNRS, Sorbonne Universit\'e, 98bis Boulevard Arago, 75014, Paris, France}
\affiliation{$^{32}$University of Ljubljana, Faculty of Mathematics and Physics, Jadranska ulica 19, SI-1000 Ljubljana, Slovenia}
\affiliation{$^{33}$Astronomical Observatory, University of Warsaw, Al. Ujazdowskie 4, 00-478 Warszawa, Poland}
\affiliation{$^{34}$Space Telescope Science Institute (STScI), 3700 San Martin Drive, Baltimore, MD 21218, USA}
\affiliation{$^{35}$Center for Astrophysical Sciences, Department of Physics and Astronomy, The Johns Hopkins University, 3400 N Charles St. Baltimore, MD 21218, USA}
\affiliation{$^{36}$Association of Universities for Research in Astronomy (AURA), Inc.~for the European Space Agency (ESA)}
\affiliation{$^{37}$Jodrell Bank Centre for Astrophysics, Alan Turing Building, University of Manchester, Oxford Road, Manchester M13 9PL, UK}
\affiliation{$^{38}$Canadian Institute for Theoretical Astrophysics, 60 St George St, University of Toronto, Toronto, ON M5S 3H8, Canada}
\affiliation{$^{39}$David A. Dunlap Department of Astronomy and Astrophysics, University of Toronto, 50 St George St, Toronto ON M5S 3H4, Canada}
\affiliation{$^{40}$Department of Physics, 60 St George St, University of Toronto, Toronto, ON M5S 3H8, Canada}
\affiliation{$^{41}$Steward Observatory, University of Arizona, 933 N Cherry Ave, Tucson, AZ 85721, USA}
\affiliation{$^{42}$David A. Dunlap Department of Astronomy \& Astrophysics, University of Toronto, 50 St. George Street, Toronto, ON M5S 3H4, Canada}
\affiliation{$^{43}$Department of Astronomy, University of Washington, Physics-Astronomy Building, Box 351580, Seattle, WA 98195-1700, USA}
\affiliation{$^{44}$eScience Institute, University of Washington, Physics-Astronomy Building, Box 351580, Seattle, WA 98195-1700, USA}
\affiliation{$^{45}$Astronomical Science Program, Graduate Institute for Advanced Studies, SOKENDAI, 2-21-1 Osawa, Mitaka, Tokyo 181-8588, Japan}
\affiliation{$^{46}$National Astronomical Observatory of Japan, 2-21-1 Osawa, Mitaka, Tokyo, 181-8588, Japan}
\affiliation{$^{47}$Observat\'orio do Valongo, Universidade Federal do Rio de Janeiro, Ladeira do Pedro Ant\^onio 43, Rio de Janeiro, RJ 20080-090, Brazil}
\affiliation{$^{48}$STAR Institute, University of Li\`ege, Quartier Agora, All\'ee du six Ao\^ut 19c, 4000 Li\`ege, Belgium}
\affiliation{$^{49}$Department of Physics, Indian Institute of Science, Bengaluru 560012, India}
\affiliation{$^{50}$MIT Kavli Institute for Astrophysics and Space Research, 70 Vassar Street, Cambridge, MA 02139, USA}
\affiliation{$^{51}$Department of Astronomy, University of Geneva, Chemin Pegasi 51, 1290 Versoix, Switzerland}
\affiliation{$^{52}$Cosmic Dawn Center (DAWN), Niels Bohr Institute, University of Copenhagen, Jagtvej 128, K\o benhavn N, DK-2200, Denmark}
\affiliation{$^{53}$National Astronomical Observatory of Japan, 2-21-1 Osawa, Mitaka, Tokyo 181-8588, Japan}
\affiliation{$^{54}$Department of Astronomical Science, SOKENDAI (The Graduate University for Advanced Studies), 2-21-1 Osawa, Mitaka, Tokyo, 181-8588, Japan}
\affiliation{$^{55}$Kavli Institute for the Physics and Mathematics of the Universe (WPI), The University of Tokyo, 5-1-5 Kashiwanoha, Kashiwa, Chiba 277-8583, Japan}
\affiliation{$^{56}$Instituto de F\'isica de Cantabria (CSIC-UC), Avda. Los Castros s/n, 39005 Santander, Spain}
\affiliation{$^{57}$Department of Physics \& Astronomy, University of California, Los Angeles, 430 Portola Plaza, Los Angeles, CA 90095, USA}
\affiliation{$^{58}$Department of Physics and Astronomy, Tufts University, 574 Boston Avenue, Suite 304, Medford, MA 02155, USA}
\affiliation{$^{59}$INAF - Observatory of Rome, Via Frascati 33, 00078, Monte Porzio Catone, Italy}
\affiliation{$^{60}$DTU Space, Technical University of Denmark, Elektrovej 327, DK-2800 Kgs. Lyngby, Denmark}
\affiliation{$^{61}$Niels Bohr Institute, University of Copenhagen, Jagtvej 155A, DK-2200 Copenhagen N, Denmark}
\affiliation{$^{62}$Department of Physics, Graduate School of Science, The University of Tokyo, 7-3-1 Hongo, Bunkyo, Tokyo 113-0033, Japan}

\footnotetext[*]{Email: seiji.fujimoto@utoronto.ca}

\begin{abstract}
We report the discovery and spectroscopic confirmation of \snh, a strongly lensed, multiply imaged Type II supernova (SN) at \zsn behind the galaxy cluster \cluster.
The transient was identified as a red F150W-dropout source (F277W $=26.1$\,mag) in \textit{JWST}/NIRCam imaging taken on 2026 July 3 for the VENUS survey, with no counterpart in 2024 December imaging of the same field.
Director's Discretionary Time observations on 2026 August 17 added second-epoch NIRCam imaging and a deep NIRSpec prism spectrum.
The spectrum shows a strong, spectrally resolved \ha emission line (FWHM $\approx7900$\,\kms) with a P-Cygni absorption blueshifted by $8800$\,\kms, together with corresponding \hb and \hei features, establishing a hydrogen-rich photosphere at $z=3.34$.
Template fits that forward-model the prism resolution classify \snh as a Type II SN, and light-curve comparisons with well-studied local SNe~II yield a phase of $27.5\pm2.8$ rest-frame days past \textit{B}-band maximum.
The 0.9\,mag drop in the rest-frame \textit{B} band over 10 rest-frame days most likely reflects post-peak photospheric evolution driven by radioactive heating, rather than a hydrogen recombination-driven plateau or its fall-off.
Independent cluster lens models magnify the observed image by $\mu\approx10$--20 and predict another image within $\simeq2$--6\,yr, offering a rare opportunity for a time-delay measurement of $H_{0}$ with a source at $z>3$.
The host galaxy is undetected in deep rest-frame ultraviolet and optical imaging, in emission lines, and in ALMA 2\,mm dust continuum, leaving a heavily obscured counterpart unlikely. Its de-lensed $M_{\rm UV}\gtrsim-13$ places it far below blank-field limits and adds support to an elevated core-collapse rate per unit star formation in ultra-faint galaxies.
Without the multi-epoch data, \snh would masquerade as a convincing $z\approx16$ Lyman-break galaxy, with galaxy-template fits strongly disfavoring any low-redshift solution even with multiple medium bands.
Its spectrum thus provides an empirical SED template with which high-redshift searches can vet similar dropout candidates.
\end{abstract}


\section{Introduction} \label{sec:intro}

Strongly lensed supernovae (SNe) with resolved multiple images provide time delays that measure the Hubble constant $H_{0}$ independently of the local distance ladder \citep[e.g.,][]{refsdal1964,kelly2015,treu2022b}.
A small sample of cluster-scale, multiply imaged SNe that are useful for cosmology is known to date, including SN Refsdal at $z=1.49$ \citep{kelly2015}, SN H0pe at $z=1.78$ \citep{frye2024}, SNe Encore and Requiem at $z=1.95$ \citep{rodney2021requiem,pierel2024encore}, SN Ares at $z=1.276$ (DD~9478; PI Larison), SN Athena at $z=0.87$ (DD~12781; PI Pascale), and SN Atalanta at $z=1.48$ (DD~12774; PI Agrawal), and each new system with a distinct lensing configuration adds an independent anchor for time-delay cosmography.
Although the value of $H_0$ is most sensitive to the deflector redshift in the system, the angular diameter distance to the source also contributes. When the distance between the lens and source is large, the system can even be used to test possible departures from the standard expansion history, such as an evolving dark energy equation of state parameter, $w$ \citep{Birrer:2024}. Such independent probes of high-redshift cosmology are critical as we enter into the era of \textit{Roman}, where SN Ia systematics, most critically the impact of dust, must be tested against independent probes such as time-delay cosmography \citep{Popovic:2026}.

High-redshift lensed systems are also valuable astrophysically, as the SNe themselves probe massive-star explosions at epochs where spectrophotometric confirmation remains scarce \citep[e.g.,][]{decoursey2025jades,siebert2024,pierel2024b,pierel_testing_2025,aparicio_analysis_2026,coulter_discovery_2026,fox_expanding_2026,yan_pearls_2026}. SN~Eos at $z=5.13$, for example, received a significant boost in signal ($\mu\approx30$) due to cluster lensing, which resulted in an incredibly high signal-to-noise spectrum that allowed for a direct measurement of metallicity in the early Universe from an individual stellar source for the first time \citep{coulter2026eos}. Another high-redshift core-collapse SN (CCSN) at $z=3$ showed a remarkable detection of shock-cooling immediately after the explosion epoch, a period of SN physics that is scarcely observed even in the local Universe and only possible at this redshift due to lensing \citep{Chen2022}.

Despite their incredible usefulness to cosmology and many fields of astrophysics, lensed SNe have historically been difficult to find. The advent of large, multi-epoch \jwst\ surveys of massive lensing clusters is opening a new discovery space for strongly lensed transients at high redshift. Among these efforts, the Cycle 4 VENUS survey (GO~6882; PIs Fujimoto \& Coe) combines deep, multi-band NIRCam imaging with repeated observations of massive lensing clusters, including fields with existing NIRCam imaging from earlier \jwst\ programs, enabling systematic searches for multiply imaged SNe and their delayed reappearances. VENUS is already uncovering remarkable strongly lensed SNe, including the aforementioned SNe Ares, Athena, Eos, and Atalanta. Here we report the discovery of \snh, identified by combining the first-epoch VENUS imaging of \cluster\ with earlier \jwst\ imaging obtained as part of the SLICE survey (GO~5594; PI Mahler), and its subsequent spectroscopic confirmation as a Type II SN at \zsn\ through Director’s Discretionary Time (DDT) observations (DD~12782; PI Fujimoto).
The combination of a hydrogen-rich SN, a massive cluster lens, and model-predicted future images makes \snh\ a promising system for a time-delay cosmology measurement. Its predicted reappearance lies years rather than decades ahead (Section~\ref{sec:lens}), and at \zsn\ it nearly doubles the redshift of the previous highest-redshift multiply imaged SNe awaiting a forecasted future image, SN Requiem and SN Encore which occurred in the same $z\simeq1.95$ host galaxy, so it exploded $1.4$\,Gyr earlier in cosmic time \citep{rodney2021requiem,pierel2024encore}.

This Letter is organized as follows.
Section~\ref{sec:obs} summarizes the observations.
Section~\ref{sec:class} presents the classification and redshift of \snh.
Section~\ref{sec:lens} presents the lens-model predictions for the magnifications and the reappearance.
Section~\ref{sec:discussion} discusses the prospects for time-delay cosmography, revisits an initial $z\approx16$ dropout interpretation of the discovery photometry, and summarizes our key results.
Throughout we adopt a flat $\Lambda$CDM cosmology with $H_{0}=70\,{\rm km\,s^{-1}\,Mpc^{-1}}$ and $\Omega_{\rm m}=0.3$, and AB magnitudes.

\section{Observations and Data Processing} 
\label{sec:obs}

Table~\ref{tab:obs} summarizes all \textit{JWST} observations used in this Letter, and Figure~\ref{fig:discovery} shows the field and the three imaging epochs at the SN position.

\subsection{JWST Data} \label{sec:jwst}

\subsubsection{Imaging} 
\label{sec:imaging}

\cluster was observed by VENUS on 2026 July 3 (MJD 61224) in ten NIRCam bands (F090W through F444W) with a four-point \texttt{INTRAMODULEBOX} dither pattern; per-filter exposure times of $1.2$--$2.1$\,ks are listed in Table~\ref{tab:obs}.
The imaging is reduced with the standard VENUS pipeline built on \texttt{grizli} \citep{brammer2023}, as in previous VENUS studies \citep[e.g.,][]{nakane2025, golubchik2026, coulter2026eos, yanagisawa2026, allingham26, asada2026eos}, which applies the level-2 detector calibrations with corrections for $1/f$ noise, snowball artifacts, and wisp features, registers the astrometry to Gaia~DR3, and drizzles all bands to a common $0\farcs03$ pixel grid.
Visual inspection of the mosaics revealed a compact source at (R.A., Decl.) $=$ (00:18:31.50, $+$16:26:18.0) that is absent in earlier imaging of the field (Section~\ref{sec:archival}).
The discovery-epoch photometry is listed in Table~\ref{tab:phot}.
The SED is red, with a $\gtrsim2$\,mag break across F150W--F200W and no clear detection blueward of F200W.

Second-epoch NIRCam imaging was obtained on 2026 August 17 (MJD 61269) under DD~12782 in F150W, F200W, F356W, and F444W, using a six-point \texttt{INTRAMODULEX} dither ($773$\,s per filter; Table~\ref{tab:obs}).
A third epoch in seven bands is scheduled for 2026 November--December.
The DDT epochs are processed with the same pipeline onto the same pixel grid, tied to the VENUS astrometry, and matched-aperture photometry of field stars agrees between the epochs to within $1\%$.

The discovery-epoch fluxes in Table~\ref{tab:phot} are total fluxes from the VENUS photometric catalog, measured and corrected in the same way as in other VENUS studies. For the epoch-to-epoch light curve we instead measure the transient with aperture photometry and a PSF correction, using $0\farcs2$-diameter apertures at the SN position, and aperture corrections to total flux from the encircled energy of empirical PSFs built from field stars. Uncertainties are rescaled to the measured scatter of blank-sky apertures. The transient is an unresolved point source, so this keeps the measurement as simple as possible and applies the correction appropriate to a point source. In the four filters covered at both epochs we measure $28.80\pm0.40$, $26.62\pm0.05$, $26.26\pm0.03$ and $26.33\pm0.04$\,mag in F150W, F200W, F356W and F444W on 2026 July 3, and $>27.6$ (2$\sigma$), $27.52\pm0.43$, $26.44\pm0.08$ and $26.49\pm0.09$\,mag on 2026 August 17.

\subsubsection{Spectroscopy} 
\label{sec:spec}

As a part of our DDT program, a NIRSpec fixed-slit spectrum of \snh was obtained on 2026 August 17 with the S200A1 aperture and the prism/CLEAR disperser, covering $0.5$--$5.5$\,$\mu$m at $R\simeq30$--$360$.
Twenty exposures of $656.5$\,s each ($13.1$\,ks on source) were taken with a 5-point nod along the slit, four subpixel dithers per nod position, and NRSIRS2 readout.
Target acquisition used a wide-aperture acquisition on a nearby star followed by a blind offset.
The data are reduced with \texttt{msaexp} \citep[v0.9.17;][]{brammer2023b} using the standard fixed-slit defaults, including nod differencing for background subtraction and optimal extraction along the fitted spatial profile.

\subsection{Pre-explosion Imaging} 
\label{sec:archival}

\begin{figure*}
\centering
\includegraphics[width=0.98\linewidth]{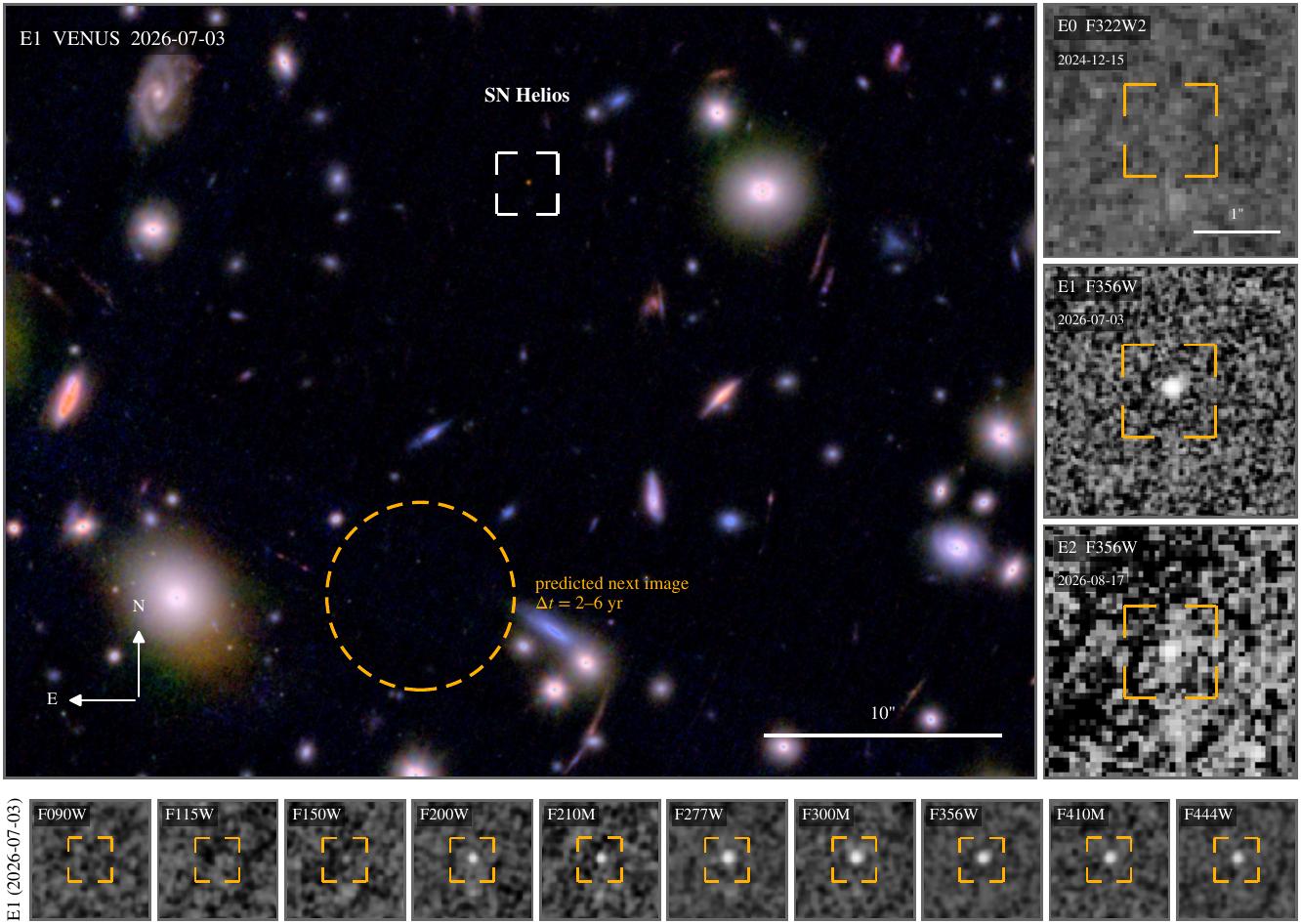}
\caption{
Discovery of \snh behind \cluster.
Top: VENUS NIRCam color image of the cluster core at the discovery epoch (2026 July 3; R $=$ F356W$+$F444W, G $=$ F200W$+$F277W, B $=$ F115W$+$F150W, $44\arcsec\times33\arcsec$).
The corner brackets mark \snh.
The dashed circle encloses the positions where the five lens models predict the next image of the SN to appear. Their predicted delays after the discovery epoch are $2.3\pm0.4$\,yr (\texttt{ALP}), $2.7\pm0.5$\,yr (Zitrin-analytic), $3.09\pm1.01$\,yr (\texttt{glafic}), $4.2\pm0.8$\,yr (\texttt{WSLAP+}) and $6.0^{+1.2}_{-0.2}$\,yr (\texttt{LENSTOOL}), placing the reappearance between 2028 and 2032. The models agree on where the next image appears to within a few arcseconds. The spread in the predicted delays is primarily due to the scarcity of spectroscopic redshifts for the multiple-image systems rather than to differences between the modeling methods (Section~\ref{sec:lens}).
Right: $3\arcsec\times3\arcsec$ cutouts at the SN position, oriented north up and east left, comparing matched wavelengths across the three epochs on an identical surface-brightness scale, each at its native pixel scale. No source is present before the explosion (E0 F322W2), the SN appears at discovery (E1 F356W), and it has faded six weeks later (E2 F356W). The second DDT epoch is intrinsically shallower ($773$\,s versus $1246$\,s), which the shared scale makes visible.
Bottom: the same cutouts in all ten VENUS bands at the discovery epoch, shown on a single common display scale, where the SN is undetected in F090W--F150W and clearly detected from F200W redward (Section~\ref{sec:redshift}).
}
\label{fig:discovery}
\end{figure*}

The field was observed with \jwst/NIRCam on 2024 December 15 in the F150W2 and F322W2 wide filters (GO 5594; PI Mahler; e.g., \citealt[][]{cerny2026}), and with \hst\ as part of RELICS \citep[e.g.,][]{coe2019}.
\snh is not detected in any pre-explosion data (Figure~\ref{fig:discovery}), which determines the explosion epoch between 2024 December and 2026 July and provides deep limits at the SN position, including a $2\sigma$ limit of F150W2 $>29.9$\,mag that also constrains the brightness of the SN host galaxy (Section~\ref{sec:lens}).

\begin{deluxetable}{lcccc}
\tablecaption{NIRCam imaging of \snh \label{tab:obs}}
\tablehead{
\colhead{Program} & \colhead{UT date} & \colhead{Filter} & \colhead{$N_{\rm exp}$} & \colhead{$t_{\rm exp}$ (s)}
}
\startdata
5594 & 2024-12-15 & F150W2 & 9 & 1836.0 \\
5594 & 2024-12-15 & F322W2 & 9 & 1836.0 \\
6882 & 2026-07-03 & F090W & 4 & 2061.5 \\
6882 & 2026-07-03 & F115W & 4 & 1674.9 \\
6882 & 2026-07-03 & F150W & 4 & 1245.5 \\
6882 & 2026-07-03 & F200W & 4 & 1245.5 \\
6882 & 2026-07-03 & F210M & 4 & 2061.5 \\
6882 & 2026-07-03 & F277W & 4 & 1245.5 \\
6882 & 2026-07-03 & F300M & 4 & 2061.5 \\
6882 & 2026-07-03 & F356W & 4 & 1245.5 \\
6882 & 2026-07-03 & F410M & 4 & 2061.5 \\
6882 & 2026-07-03 & F444W & 4 & 1674.9 \\
12782 & 2026-08-17 & F150W & 6 & 773.0 \\
12782 & 2026-08-17 & F200W & 6 & 773.0 \\
12782 & 2026-08-17 & F356W & 6 & 773.0 \\
12782 & 2026-08-17 & F444W & 6 & 773.0 
\enddata
\end{deluxetable}

\begin{deluxetable}{lcc}
\tablecaption{Discovery-epoch photometry of \snh (2026 July 3) \label{tab:phot}}
\tablehead{
\colhead{Filter} & \colhead{$f_{\nu}$ (nJy)} & \colhead{AB mag}
}
\startdata
F090W & $-2.3\pm4.4$ & $>29.0$ \\
F115W & $3.0\pm4.7$ & $>29.0$ \\
F150W & $6.5\pm5.1$ & $>28.9$ \\
F200W & $61.6\pm4.3$ & $26.93\pm0.08$ \\
F210M & $54.3\pm5.0$ & $27.06\pm0.10$ \\
F277W & $128.9\pm3.0$ & $26.12\pm0.03$ \\
F300M & $110.2\pm3.3$ & $26.29\pm0.03$ \\
F356W & $101.1\pm2.8$ & $26.39\pm0.03$ \\
F410M & $100.5\pm4.0$ & $26.39\pm0.04$ \\
F444W & $84.2\pm3.3$ & $26.59\pm0.04$ 
\enddata
\tablecomments{
Total fluxes from the VENUS photometric catalog.
Bands with flux below $2\sigma$ are quoted with their $2\sigma$ upper limits. }
\end{deluxetable}

\section{Classifying SN Helios} \label{sec:class}

\subsection{Redshift} \label{sec:redshift}

\begin{figure*}
\centering
\includegraphics[width=0.95\linewidth]{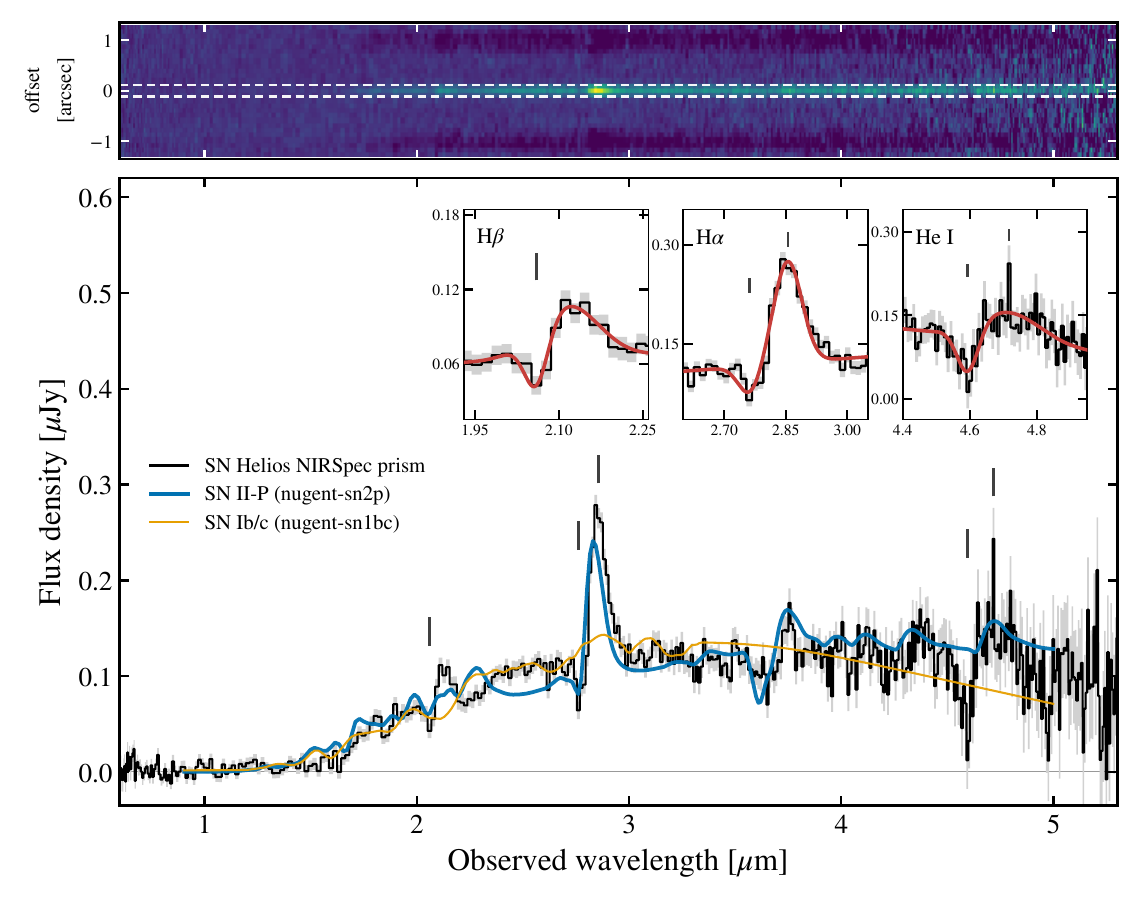}
\caption{
\jwst/NIRSpec S200A1 prism spectrum of \snh ($13.1$\,ks, 2026 August 17; black with gray $1\sigma$ band), with the best-fit Type IIP template (blue) and, for contrast, the best-fit stripped-envelope Ib/c template (orange), both forward-modeled through the wavelength-dependent prism line-spread function with the phase, host reddening, and amplitude free.
The top panel shows the rectified two-dimensional spectrum; dashed lines mark the FWHM ($0\farcs23$) of the fitted spatial profile used for the optimal extraction. 
The insets zoom on the \hb, \ha, and \hei features. The red curves show the best-fit P-Cygni profile with a linear continuum plus emission and absorption Gaussians.
The hydrogen-rich templates are preferred over the stripped-envelope and thermonuclear alternatives at $\Delta\chi^{2}$ of several hundred, driven by the broad \ha emission that the hydrogen-poor templates cannot produce.
}
\label{fig:classification}
\end{figure*}

Figure~\ref{fig:classification} presents the prism spectrum, which shows a red continuum ($\simeq0.11\,\mu$Jy, $26.4$\,mag at $3$--$5$\,$\mu$m) with a strong emission line at $2.855$\,$\mu$m detected at ${\rm S/N}\simeq24$.
Interpreting the line as \ha places the emission peak at $z=3.350$; for a broad P-Cygni profile the emission peak does not generally coincide with the systemic wavelength.
An alternative identification as \oiii\,$\lambda5007$ at $z=4.70$ is disfavored on several grounds. The line is spectrally resolved and carries a blueshifted absorption trough (Section~\ref{sec:spec}), whereas \oiii\ is a forbidden transition from low-density gas that produces neither a $\sim8000$\,\kms width nor a P-Cygni profile. At $z=4.70$ \hb would fall at $2.77$\,$\mu$m, where the spectrum instead shows a deep absorption trough, so the implied \oiii\,$\lambda5007$/\hb and \ha/\hb ratios are negative and no H\,{\sc ii} region can reproduce them. The weak feature near $3.76$\,$\mu$m that would then be \ha is instead matched by the Ca\,{\sc ii} near-infrared triplet at \zsn.
We also perform an independent, empirical cross-check by fitting the prism spectrum of SN Eos at $z=5.13$ \citep[DD~9493;][]{coulter2026eos,asada2026eos}, the only CCSN with a high signal-to-noise prism spectrum at a comparable redshift, so that the rest-frame coverage and the wavelength-dependent line-spread function match those of \snh, as a template with only redshift and normalization free; the best match is at $z=3.338$ (Appendix~\ref{app:eos}).
We adopt \zsn throughout this paper, and all line velocities are quoted relative to the systemic wavelengths at this redshift.
Because the redshift derives from broad SN lines rather than narrow host lines, we evaluate a systematic uncertainty of $\Delta z\simeq0.01$.

\subsection{A hydrogen-rich photosphere} 
\label{sec:hydrogen}

Line features are measured on the 1D spectrum as integrated fluxes against linear continuum fit to flanking windows, with significances propagated from the error spectrum; absorption velocities are measured by fitting a Gaussian to each trough over a local linear pseudo-continuum, in the same manner as \citet{Gutierrez2017}, and are quoted relative to the systemic wavelengths at \zsn.

The $2.855$\,$\mu$m line feature is spectrally resolved.
The observed FWHM of $810$\,\AA\ compares to the instrumental resolution of $310$\,\AA\ at this wavelength ($R=92$), giving an intrinsic ${\rm FWHM}\approx7900$\,\kms.
A P-Cygni absorption trough is robustly detected at $2.761$\,$\mu$m with the $5.5\sigma$ level, blueshifted from systemic \ha by $8800\pm500$\,\kms.
The corresponding \hb absorption is also present at $2.060$\,$\mu$m, blueshifted by $7100\pm1500$\,\kms, and a \hei P-Cygni profile is further detected at $4.59/4.72$\,$\mu$m, blueshifted by $6800\pm500$\,\kms.
Within the prism resolution and its lower signal-to-noise, \hb\ shows the same P-Cygni profile as \ha, with the two absorption velocities separated by $1600\pm1600$\,\kms, consistent with the \ha--\hb\ velocity hierarchy of local SNe~II at this phase \citep[$\simeq700$--$900$\,\kms\ at $+23$--$28$\,d;][]{Gutierrez2017}.
These features describe a single photosphere expanding at $7$--$9\times10^{3}$\,\kms with a substantial hydrogen envelope, which establishes \snh as a Type II SN independent of any template fitting and which is consistent with the local sample of SNe II at earlier phases \citep[see Sec.~\ref{sec:phase} for a more detailed discussion,][]{Gutierrez2017}.

\begin{figure}
\centering
\includegraphics[width=\linewidth]{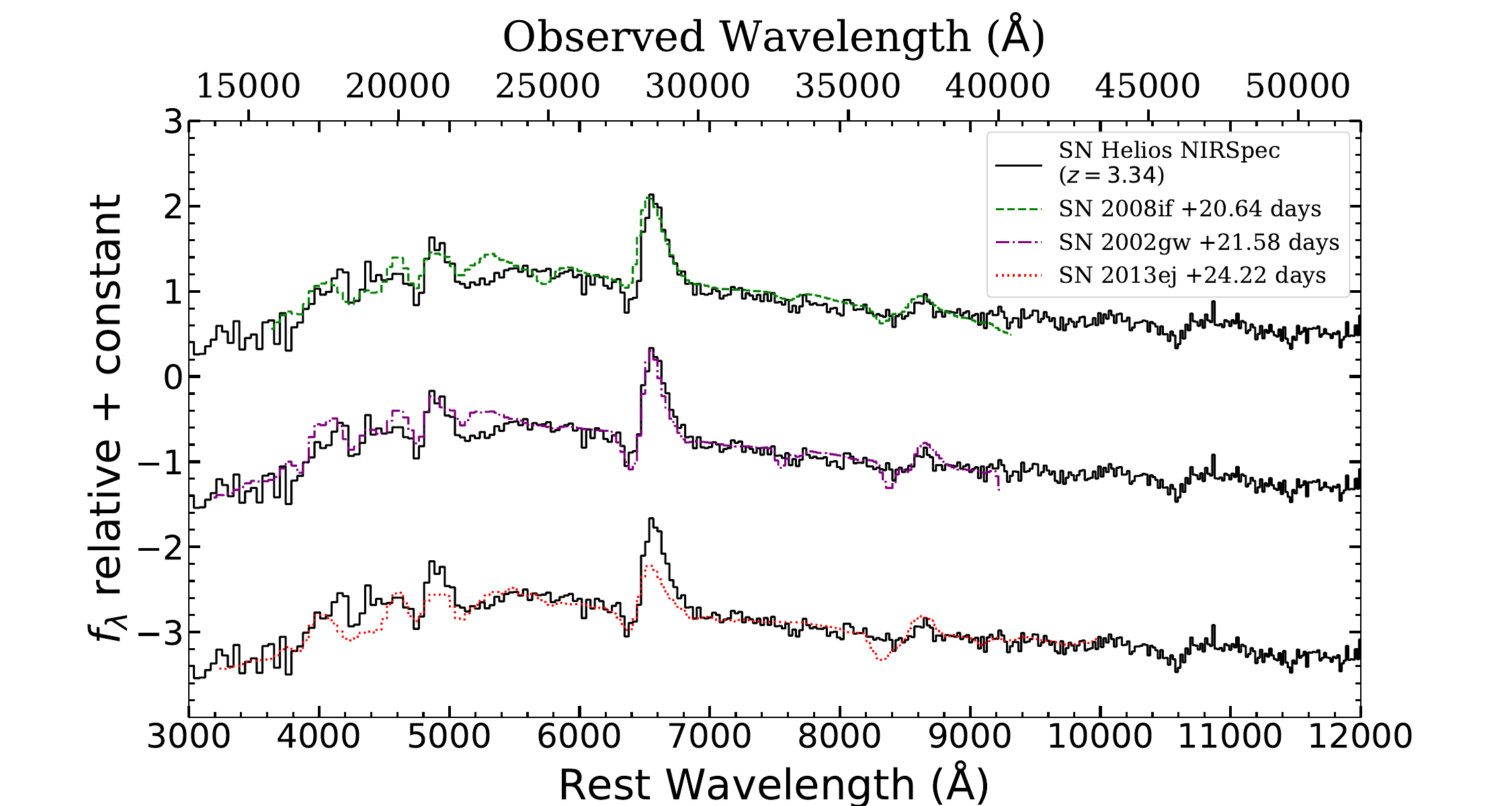}
\caption{Spectroscopic comparisons between \snh and three SNe II at a phase of $\sim+20 - 25$ days post \textit{B}-band maximum. Comparisons include the fast-declining SN II, SN 2013ej (a light curve comparison to which is shown in Figure~\ref{fig:lightcurve}) and two SNe II from the sample presented by \citet{Gutierrez2017}. SN 2002gw and SN 2008if were selected for having low $\chi^2_{\nu}$ values when matching the continuum-subtracted \ha feature; however, it is clear that their continuua shapes also match closely to \snh at this phase. The close continuum match to SN~2013ej is expected given its similar light curve evolution to \snh (see Section~\ref{sec:lightcurve}).
}
\label{fig:specfit}
\end{figure}

\subsection{Spectral classification and phase} 
\label{sec:phase}

The template comparison in Figure~\ref{fig:classification} forward-models each SN template through the wavelength-dependent prism line-spread function, with the phase, host reddening, and amplitude free. SNID \citep{blondin2007} comparisons of the de-redshifted spectrum against local SN templates confirm the classification, with SN 1999em and SN 2005cs the closest individual matches. These comparisons are coarse; however, as SNID lacks the temporal coverage and exact matched prism resolution required to recover a robust estimate of phase. To gain a better estimate, we compare to the large sample of SN II spectra presented by the Carnegie Supernova Project \citep[CSP, ][]{Gutierrez2017}. In Figure~\ref{fig:specfit}, we compare \snh\ with three SNe~II observed $\sim20$--$25$ days past \textit{B}-band maximum, which reproduce both its continuum shape and its \ha\ profile. We specifically examine differences in the \ha P-Cygni profile, which is resolved for \snh, as it has been shown that both the FWHM and velocity shift of this line correlates with phase \citep{Gutierrez2017}. To do this, we isolate the feature for both \snh and the comparison object to measure a $\chi^2_\nu$ value, with the total uncertainty assumed to be from the \snh spectrum. We interpolate the SN II comparison spectrum so that it is at the same prism-resolution scale as \snh and then fit and subtract the continuum around the \ha feature, taking the bounds to be 6250 \AA\ and 6900 \AA\ for the blue and red limits, respectively. It is important to match the resolution of the comparison spectra, rather than simply measuring the FWHM and emission/absorption line velocity shifts directly and comparing to low-redshift values, as the prism resolution may introduce systematic differences in recovered values that would bias our phase estimate.

The results of this procedure are explained in more detail in the Appendix (see Figure~\ref{fig:csp_spec}). It is clear that there are a wide range of phases at which a low $\chi^2_\nu$ value is recoverable, although there is a clear clustering of low $\chi^2_\nu$ values in the $20 - 30$ day phase range. Due to the large scatter in values, it is difficult to ascertain an accurate uncertainty from this analysis, but the evidence from this comparison of \ha alone points strongly to this phase range for \snh. Further spectroscopic observations of \snh when it reappears will allow for a re-analysis with more spectrosopic coverage and further permit a direct spectroscopic time-delay estimate as has been done for other strongly lensed SNe \citep{Johansson:2021,Chen:2024,Johansson:2026}. Given this phase range, our spectrum of \snh\ is too early to perform a similar Fe II line analysis as was done with SN~Eos to infer the gas-phase metallicity of its host environment \citep{coulter2026eos}. Additionally, we lack the spectral resolution to resolve the Fe II 5018 and 5169 \AA\ lines in the triplet; however, a reappearance of \snh\ could allow for this more in-depth analysis to be done in conjunction with fitting for a spectrophotometric time delay.

\subsection{Photometric evolution} 
\label{sec:lightcurve}

The two NIRCam epochs are separated by 45 observer-frame days ($10.4$ rest-frame days at \zsn); the photometric measurements are described in Section~\ref{sec:imaging}.
\snh fades by $0.90\pm0.58$\,mag in F200W (rest-frame $B$ band) between the two epochs, while declining only mildly in F356W and F444W (rest-frame $I$ and $Y$; $0.18\pm0.13$ and $0.16\pm0.12$\,mag).

The nearly flat rest-frame $I$-band evolution is roughly the behavior expected of a Type II in the redder bands; however, the relatively fast drop in the rest-frame \textit{B}-band implies that the optical plateau will begin at a later phase that may still be in progress at the scheduled 2026 November--December third epoch. This is consistent with our spectroscopic analysis of \snh, which also favors a pre-plateau phase.

To estimate the phase of \snh based on its light curve, we rely on well-sampled local SNe II for which rest-frame $U$ through $J$ observations are available. As opposed to Type Ia supernovae which have a homogeneous population in terms of light curve evolution, SNe II have a large diversity that makes it intractable to construct reliable light curve templates for the population as an aggregate \citep{Hillier:2019}. We therefore rely on finding the best-matched analog to \snh\ from which we can attempt to estimate the phase. We do this by employing a method comparing Gaussian process interpolations of well-sampled ground-based data to the \snh photometry. Because these commonly used ground-based filters do not match the response functions of the \textit{JWST} filters, we first correct for the differences in flux caused by this mismatch. We do so by sampling synthetic photometry of the available spectra across time, and we use a simple linear interpolation between phases for which no spectra were available. We use all available spectra from the Open Supernova Catalog (OSC) for our analysis \citep[references to sources cited in Table~\ref{tab:template_data},][]{Guillochon2017}. We redshift the spectroscopic observations of the comparison objects to the observed frame and correct for Milky Way dust extinction to then recover synthetic photometry with the \textit{JWST} response functions. Milky Way reddening values are from the \citet{Schlafly:2011} recalibration of the \citet{Schlegel:1998} dust maps. The dust corrections are applied assuming an $R_V$ of 3.1 and the \citet{Fitzpatrick:1999} dust model. We apply the shift in these values from the ground-based filters to the comparison SN photometry to accurately match the photometric evolution to \snh. A more detailed explanation of this methodology will be presented in Larison et al. (in prep.). For this comparison, we apply this light curve fitting method to three well-studied local SNe II that are densely sampled in terms of both phase and wavelength: SN 1999em, SN 2005cs, and SN 2013ej. A summary of the assumptions that go into their respective light curve fits are supplied in Table~\ref{tab:template_assumptions}. From this approach, we recover phase estimates from the three SNe of $30.3\pm5.2$ days, $26.4\pm4.6$ days, and $26.1\pm5.0$ days of the second \textit{JWST} epoch relative to \textit{B}-band maximum, respectively. All three estimates agree quite well to each other, within $1\sigma$, and all indicate that \snh is still pre-plateau during the second epoch of observations. Further, these phase estimates strongly agree with the $20-30$ day phase estimate obtained with the \ha feature analysis. Taking the weighted average of these results, we recover a phase estimate of $27.5 \pm 2.8$ days relative to \textit{B}-band maximum.

\begin{table*}
\centering
\caption{Photometric and spectroscopic data used for the Type II
supernova template analyses. The photometric counts include only the
bands used to construct the four template light curves for the SN Helios
fit. References are listed below the table.}
\label{tab:template_data}

\begin{tabular}{lccp{3.3cm}cp{3.7cm}}
\hline\hline
SN &
Photometric bands &
$N_{\rm phot}$ &
Photometric references &
$N_{\rm spec}$ &
Spectroscopic references \\
\hline

SN 1999em &
$U(30),\,B(35),\,I(35),\,Z(17)$ &
117 &
\citetalias{Galbany2016} &
101 &
\citetalias{Leonard2002},
\citetalias{Hamuy2001},
\citetalias{Faran2014},
\citetalias{Silverman2012},
\citetalias{Gutierrez2017},
\citetalias{YaronGalYam2012} \\

SN 2005cs &
$U(34),\,B(123),\,I(101),\,J(21)$ &
279 &
\citetalias{Pastorello2009},
\citetalias{Faran2014},
\citetalias{Brown2014},
\citetalias{Dessart2008},
\citetalias{TsvetkovPavlyuk2014} &
18 &
\citetalias{Pastorello2009} \\

SN 2013ej &
$u(24),\,B(182),\,I(118),\,Y(27)$ &
351 &
\citetalias{Yuan2016},
\citetalias{Huang2015},
\citetalias{Brown2014},
\citetalias{Dhungana2016} &
122 &
\citetalias{Yuan2016},
\citetalias{Valenti2014},
\citetalias{Silverman2012},
\citetalias{Childress2016},
\citetalias{Dhungana2016},
\citetalias{YaronGalYam2012} \\

\hline
\end{tabular}

\vspace{0.5em}

\begin{minipage}{0.98\textwidth}
\footnotesize
\textit{References.} ---
\citetalias{Galbany2016} Galbany et al. (2016);
\citetalias{Pastorello2009} Pastorello et al. (2009);
\citetalias{Faran2014} Faran et al. (2014);
\citetalias{Brown2014} Brown et al. (2014);
\citetalias{Dessart2008} Dessart et al. (2008);
\citetalias{TsvetkovPavlyuk2014} Tsvetkov \& Pavlyuk (2014);
\citetalias{Yuan2016} Yuan et al. (2016);
\citetalias{Huang2015} Huang et al. (2015);
\citetalias{Dhungana2016} Dhungana et al. (2016);
\citetalias{Leonard2002} Leonard et al. (2002);
\citetalias{Hamuy2001} Hamuy et al. (2001);
\citetalias{Silverman2012} Silverman et al. (2012);
\citetalias{Gutierrez2017} Guti\'errez et al. (2017);
\citetalias{YaronGalYam2012} Yaron \& Gal-Yam (2012);
\citetalias{Valenti2014} Valenti et al. (2014);
\citetalias{Childress2016} Childress et al. (2016).
\end{minipage}

\end{table*}

\begin{table}
\centering
\caption{Assumptions adopted for the Type II supernova template
light-curve fits. For dereddening, we assume $R_V=3.1$ and the \citet{Fitzpatrick:1999} dust model and for the maximum MJD, we assume an uncertainty of 1 day on each, explicitly adding this uncertainty in quadrature with our final phase estimates.}
\label{tab:template_assumptions}

\begin{tabular}{lccc}
\hline\hline
SN &
$z$ &
$E(B-V)_{\rm MW}$ &
$T_{B,\mathrm{max}}$ (MJD) \\
\hline

SN 1999em &
0.002392 &
0.0346 &
51483.43 \\

SN 2005cs &
0.001370 &
0.0314 &
53553.21 \\

SN 2013ej &
0.002192 &
0.0597 &
56506.73 \\

\hline
\end{tabular}
\end{table}

In terms of a single best-match analog to the light curve evolution of \snh, we currently favor the faster-evolving SN~2013ej. For SN~1999em and SN~2005cs, the F356W and F444W light curves would predict a slight rise in the light curve for \snh between epochs, while SN~2013ej shows a decrease which matches well to our observations. This faster decline might suggest that \snh is a lower-energy explosion and could have a relatively short plateau when compared to most SNe~IIP. Further observations of both this and future images will allow us to better constrain the true phase and light curve evolution of \snh.

\begin{figure}
\centering
\includegraphics[width=\linewidth]{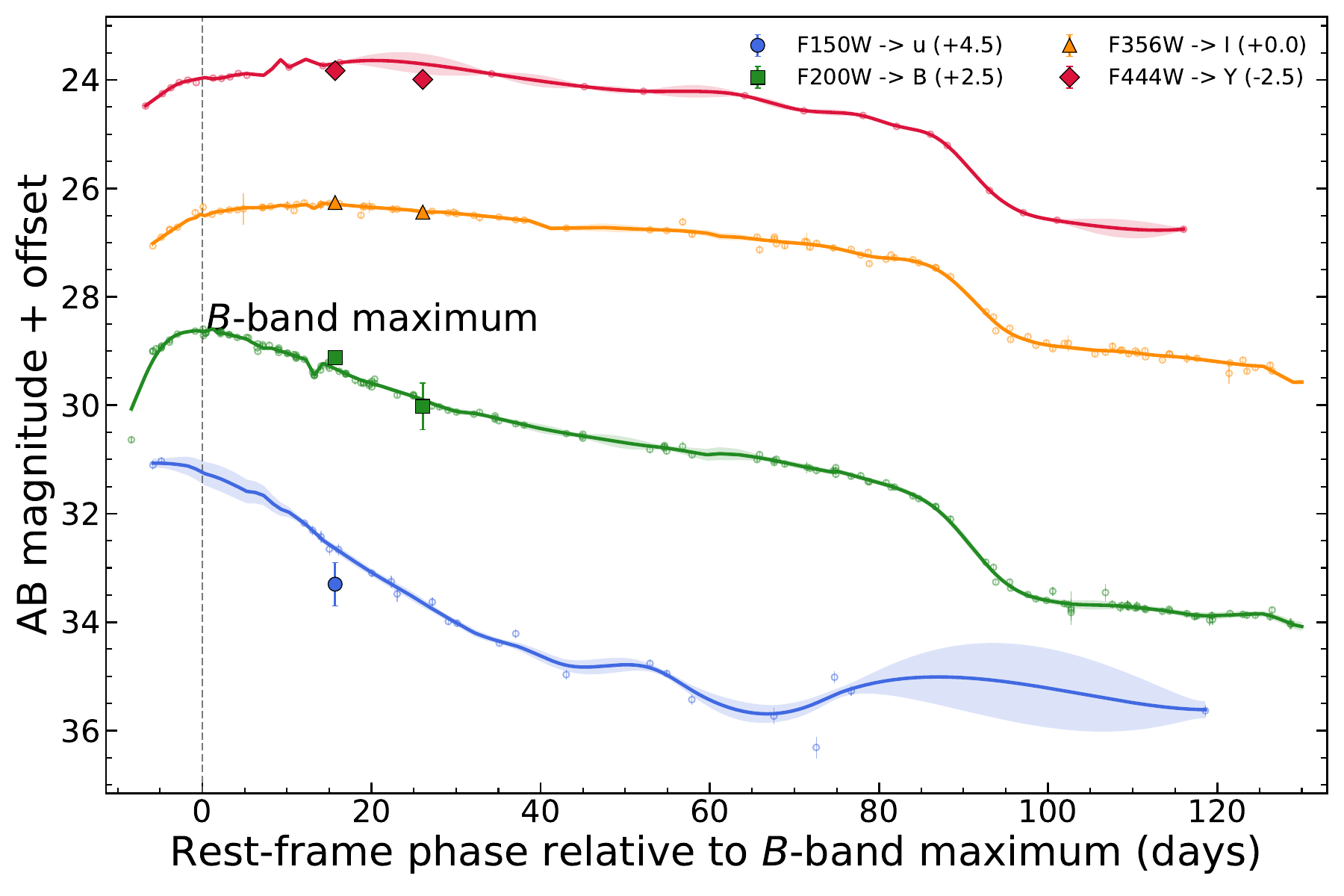}
\caption{
Type II template light-curve fit to the two-epoch NIRCam photometry, obtained by comparing against the fast-evolving SN~2013ej. We note that the shape of the F356W and F444W light curves are better matched to SN~2013ej than either SN 1999em or SN 2005cs, against which we also made comparisons (see Appendix, Figure~\ref{fig:other_lcs}). The notable features of \snh are its relatively fast decline in \textit{B}-band and its lack of a rise in F356W and F444W, which may be seen in other SNe II. Further observations will help constrain the long-term evolution of the light curve and place stronger constraints on phase.
}
\label{fig:lightcurve}
\end{figure}

\section{A Multiply Imaged Supernova at $z=3.34$} \label{sec:lens}

Five independent cluster lens models predict that \snh is multiply imaged, and identify the observed image as neither the first nor the last arrival.
Table~\ref{tab:lens} summarizes the five models and their predictions for the next image to arrive. Only a few of the multiple-image systems have measured spectroscopic redshifts, so every model is constrained almost entirely by photometrically identified systems, and this scarcity is the primary reason the predicted delays described below span nearly a factor of three. The models also differ in which images they adopt as constraints, by design as much as by circumstance. The \texttt{glafic} and \texttt{WSLAP+} models share one constraint set so that a parametric and a free-form algorithm can be compared directly, whereas the \texttt{LENSTOOL} constraints were identified independently from the SLICE and archival \hst\ imaging, which tests how much the predictions also depend on the input catalog itself.

A parametric model by L.~Furtak et al.\ (in prep.) is constructed with \texttt{AstroLensPy} \citep[\texttt{ALP};][]{allingham26}, a python implementation of the \citet{zitrin2015} method (sometimes called \texttt{zitrin-analytic}, which was natively built in \texttt{MATLAB}). The \texttt{ALP} model is constrained with 47 multiple images of 18 sources, 3 of which have spectroscopic redshifts \citep{furtak22,mainali2019}, and achieves a lens-plane average image reproduction error of $\Delta_{\rm RMS}=0\farcs30$. It predicts a five-image system for \snh. In this model the observed image is magnified by $\mu=11.2\pm0.3$, a past image with $\mu\simeq3.4$ arrived $\approx46$\,yr ago, the next image ($\mu=7.2\pm0.3$) appears $+2.3\pm0.4$\,yr after the discovery epoch (mid 2028 to early 2029) at (R.A., Decl.) $=$ (00:18:31.54, $+$16:26:00.27), and a de-magnified central pair follows on a $\approx20$\,yr delay.

To probe the range of time delays allowed by this parametrization, we also build a suite of models with the native \texttt{zitrin-analytic} pipeline \citep{zitrin2015}, constrained using the same list of cluster members as above (L.~Furtak et al.\ in prep.). To minimize possible noise from less secure systems, the basic model presented here is constrained using only 15 images of 6 background sources, 2 of which have spectroscopic redshifts, and reaches an image reproduction error of $\Delta_{\rm RMS}=0\farcs30$. Three images of the SN are predicted in total, including the currently observed one. In this model the first SN image appeared $\sim52$\,yr ago, and the next (and last) image is predicted to appear $2.70\pm0.54$\,yr after the observed SN image (Table~\ref{tab:lens}). The 10--20 other models constructed in the process, using between 6 and 12 secure image systems from the \texttt{ALP} list, one or two dark-matter halos, and different priors, all predict the same three-image configuration, with the next image spanning delays of $\sim2$--$5$\,yr after the observed image.

The independent \texttt{glafic} \citep{oguri2010,oguri2021b} model, which uses 38 multiple images from 15 systems with 2 systems having spectroscopic redshifts \citep{furtak22}, predicts three images, with the observed image at $\mu=11.5\pm3.1$ and the next image appearing $3.09\pm1.01$\,yr from now at (R.A., Decl.) $=$ (00:18:31.78, $+$16:25:59.3) with $\mu=10.9\pm3.1$. This model achieves $\Delta_{\rm RMS}=0\farcs58$, the root-mean-square of the positional differences between the observed and model-predicted multiple images (M.~Oguri et al.\ in prep.).

Using the exact same constraints as in the \texttt{glafic} model, we derive an alternative model using the hybrid method \texttt{WSLAP+} \citep{Diego2005,Diego2007,Diego2026}. This model makes no assumptions about the distribution of dark matter, while it assumes the member galaxies contain mass that is proportional to the amount of observed light. The smooth mass from the dark-matter halo is modeled as a superposition of Gaussian functions placed on a regular grid of $20\times20$ grid points covering an area of $2.'4\times2.'4$. This model predicts that \snh is magnified by $\mu=25.1$ (posterior mean $30.6\pm4.1$) and arrived $19.2$\,yr after the first image ($\mu=5.4$). The next image arrives $5.24$\,yr after \snh (posterior mean $4.18\pm0.82$\,yr, Table~\ref{tab:lens}), and two additional images are predicted $5.4$ and $7.1$\,yr after \snh with $\mu=1.3$ and $3.6$, respectively.

Finally, we computed a parametric lens model with \texttt{Lenstool} \citep{jullo2007}, constrained by 25 multiple images of 10 sources identified primarily in the NIRCam F150W2 and F322W2 imaging from the SLICE program and archival \hst\ data from RELICS. These lensing constraints were identified completely independently from those of the other models described in this paper, thus providing a complementary assessment of systematic uncertainties to the different algorithms that used the exact same constraints. Similar to the \texttt{ALP} model, we used the spectroscopic redshifts of three sources from \citet{furtak22} and \citet{mainali2019}, and all the other redshifts were left as free parameters with very broad priors, i.e., photometric redshifts did not influence the priors. The model achieves $\Delta_{\rm RMS}=0\farcs47$ and predicts a three-image configuration for the lensed SN, in which the observed image is magnified by $\mu=5.3^{+0.2}_{-0.4}$, the next image arrives $6.0^{+1.2}_{-0.2}$\,yr after it (Table~\ref{tab:lens}), and a past image with $\mu=2.4$ arrived $42.7$\,yr ago at (R.A., Decl.) $=$ (00:18:34.14, $+$16:25:39.9).

\begin{table*}
\centering
\caption{Cluster lens-model predictions for the reappearance of \snh.
``Constraints'' lists the number of multiple images and sources used by
each model (see text).
$\Delta_{\rm RMS}$ is the image-plane reproduction error and $N_{\rm img}$
the total number of SN images the model predicts. The \snh column gives the
magnification of the observed image, and the last three columns give the
predicted position, the delay relative to the observed image, and the
magnification of the next image to arrive. Uncertainties are statistical (68\%). For \texttt{WSLAP+}, the values are posterior means and standard deviations.
}
\label{tab:lens}
\begin{tabular}{lccccccc}
\hline\hline
Model &
Constraints &
$\Delta_{\rm RMS}$ &
$N_{\rm img}$ &
\snh &
\multicolumn{3}{c}{Next image} \\
 & (images/sources) & (arcsec) & & $\mu$ & (R.A., Decl.) & $\Delta t$ (yr) & $\mu$ \\
\hline
\texttt{ALP}             & 47/18 & 0.30     & 5 & $11.2\pm0.3$        & 00:18:31.54, $+$16:26:00.3 & $+2.32\pm0.44$ & $7.2\pm0.3$ \\
\texttt{zitrin-analytic} & 15/6  & 0.30     & 3 & $6.1^{+0.6}_{-0.5}$ & 00:18:31.84, $+$16:26:04.3 & $+2.70\pm0.54$ & $4.9^{+0.7}_{-0.5}$ \\
\texttt{glafic}          & 38/15 & 0.58     & 3 & $11.5\pm3.1$        & 00:18:31.78, $+$16:25:59.3 & $+3.09\pm1.01$ & $10.9\pm3.1$ \\
\texttt{WSLAP+}          & 38/15 & 1.49     & 5 & $30.6\pm4.1$              & 00:18:31.96, $+$16:25:57.0 & $+4.18\pm0.82$        & $25.1\pm10.4$ \\
\texttt{LENSTOOL}        & 25/10 & 0.47     & 3 & $5.3^{+0.2}_{-0.4}$ & 00:18:31.73, $+$16:25:58.2 & $+6.0^{+1.2}_{-0.2}$ & $3.1^{+0.1}_{-0.3}$ \\
\hline
\end{tabular}
\end{table*}

Figure~\ref{fig:discovery} marks the region enclosing the predicted positions, which agree to within a few arcseconds, while the predicted delays span $\simeq2$--$6$\,yr.
Throughout this paper we adopt as the fiducial magnification the median of the five model values, $\mu=11.2$, which also coincides with the \texttt{ALP} best fit. Every de-lensed quantity in Section~\ref{sec:discussion} scales as $1/\mu$ and is reported at this fiducial value. The models will be updated with redshifts from scheduled \textit{Keck}/KCWI observations (L.~Furtak et al.\ in prep.) well before the predicted reappearance window of \snh.

The SN host galaxy is not detected in any imaging, with a $2\sigma$ limit of F150W2 $>29.9$\,mag at the SN position (Figure~\ref{fig:host}, top left); we discuss the implied host properties in Section~\ref{sec:host}.
The absence of a detectable host means the lens models cannot use host counter-images as constraints, and the SN images themselves will become the anchoring constraints once the reappearance is observed.

\section{Discussion and Summary} \label{sec:discussion}

\subsection{Prospects for time-delay cosmography at $z>3$} \label{sec:prospects}

\snh\ is a spectroscopically confirmed, multiply imaged Type II SN at \zsn, observed near peak with a hydrogen photosphere expanding at $7$--$9\times10^{3}$\,\kms. It is among the highest-redshift spectroscopically confirmed SNe of any kind and, among multiply imaged SNe, is second in redshift only to SN Eos at $z=5.13$ \citep[e.g.,][]{coulter2026eos}. Crucially, however, \snh\ is the first spectroscopically confirmed multiply imaged SN at $z>2$ with a future image predicted to arrive on a measurable timescale, opening time-delay cosmography with SNe to a previously unexplored redshift regime.

Three properties make this system a promising cosmological tool.
First, the predicted reappearance on a $\simeq2$--$6$\,yr timescale is long enough in the future to plan for, yet a short enough window to monitor. 
Publishing the model forecasts before the reappearance follows the precedent set for SN Refsdal and SN Requiem, whose predicted fourth image (``in the year 2037$\pm$2'') was announced at discovery \citep{rodney2021requiem}, and turns the reappearance itself into a falsifiable test of the lens models.
Second, a Type IIP light curve provides a sharp clock.
The plateau-to-drop off transition occurs over a few rest-frame weeks, which at \zsn stretches to a few observer-frame months, and catching this feature in both the current and the future images would substantially tighten the delay measurement relative to a featureless light curve.
Third, SNe IIP allow for independent luminosity-distance estimates through the expanding-photosphere and standardizable-candle methods \citep{kirshner1974,hamuy2002}, which can constrain the absolute magnification of the SN and thereby address the mass-sheet degeneracy that limits purely geometric time-delay constraints.
Fortunately, the immediate next steps are in hand. The scheduled third NIRCam epoch in 2026 November--December will catch the plateau or its end, and the upcoming {\it Keck}/KCWI spectroscopy of the cluster core will secure the multiple-image redshifts that currently limit all five lens models. Additional deep imaging can pursue the host at the SN and counter-image positions. Given the current dataset for this image and the already auspicious uncertainty on its phase on the order of a few days, we estimate that a successful reappearance campaign could eventually reach an inference on $H_0$ competitive to the $2-4\%$ achieved for the current cluster lensed SNe Refsdal, H0pe, Encore, and Requiem, once the scheduled spectroscopy secures the multiple-image redshifts and the lens models converge.

In addition to the cosmological benefits, this reappearance will unlock possible explorations of SN physics that are not possible with the current dataset. For instance, a higher-resolution spectrum taken during the optical plateau would allow us to use the Fe II line complex around 5018 \AA\ as a gas-phase metallicity tracer for the SN environment as was done with SN~Eos \citep{coulter2026eos}, which would also inform our understanding of the undetected host. Further, we will be able to obtain a spectroscopic time series of \snh\ if we can detect it early in its evolution, which would make it the first $z>2$ CCSN to date with multi-phase spectroscopic observations that allow for a more in-depth understanding of explosion properties.

\subsection{The missing host galaxy} 
\label{sec:host}

\begin{figure}
\centering
\includegraphics[width=\linewidth]{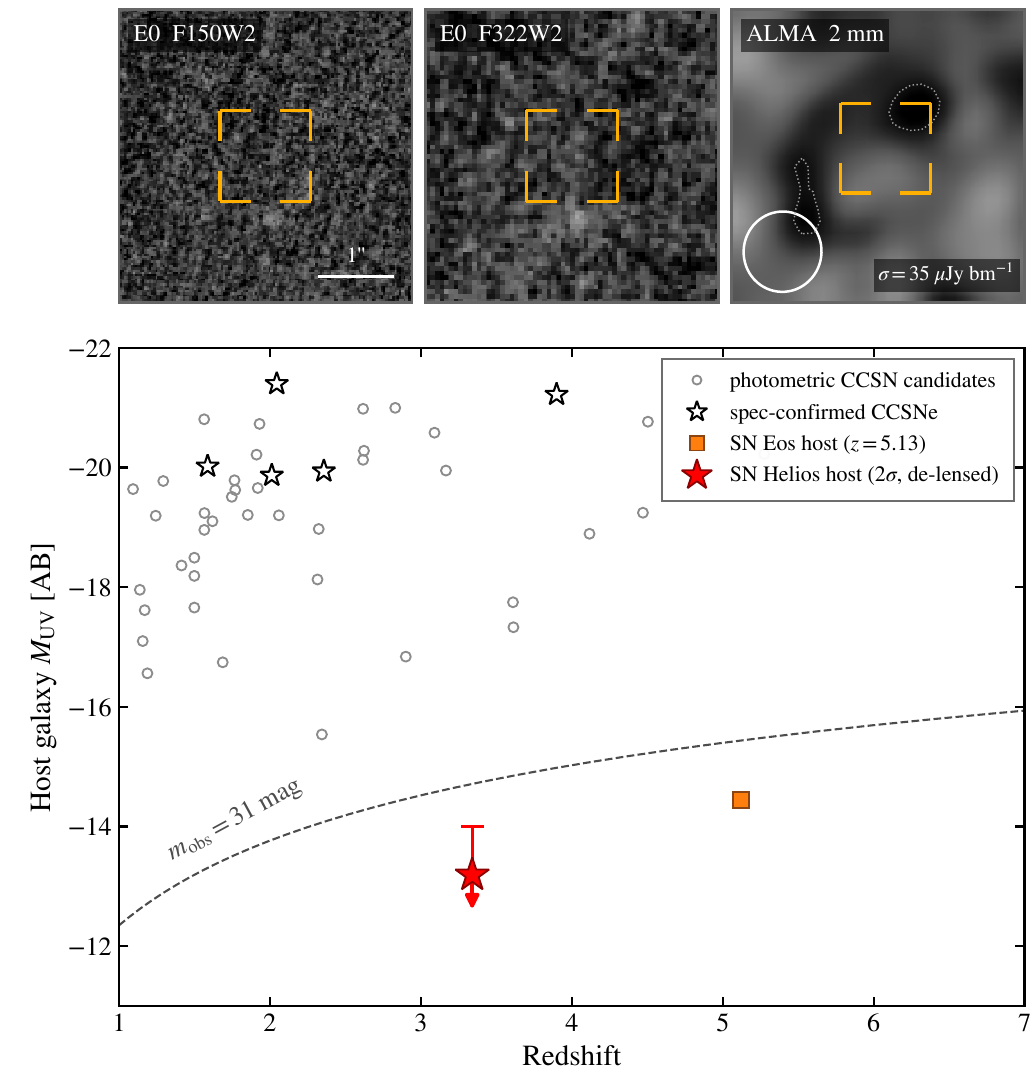}
\caption{
The \snh host in the $M_{\rm UV}$--redshift plane of CCSN hosts.
Open stars are hosts of spectroscopically-confirmed CCSNe \citep{cooke2009,cooke2012,schulze2018} and open circles hosts of photometric CCSN candidates from \jwst\ surveys \citep{decoursey2025var,decoursey2025jades,coulter2026z36}; the orange square is the SN Eos host.
Comparison points were taken from the compilation in \citet{asada2026eos}.
The red star is the de-lensed $2\sigma$ upper limit on the \snh host ($\mu=11.2$; the tick marks the lowest model value, $\mu=5.3$), converted from rest-frame $3800$\,\AA\ assuming a flat UV slope.
The dashed curve marks $M_{\rm UV}$ for an observed magnitude of 31, the depth frontier of \jwst\ imaging; without the lensing magnification the \snh host would be invisible even there.
The top row shows $4\arcsec\times4\arcsec$ cutouts at the SN position (north up, east left). The pre-explosion F150W2 and F322W2 imaging sets the host limit plotted below, and the ALMA 2\,mm map (M.~VanWyngarden et al. in prep.; contours at $\pm2$ and $\pm3\sigma$, beam at lower left) shows no dust continuum, so the host escapes detection from the rest-frame ultraviolet to the millimeter.}
\label{fig:host}
\end{figure}

The deep pre-explosion F150W2 imaging places a $2\sigma$ limit of $>29.9$\,mag at the SN position (Figure~\ref{fig:host}, top left), which after removing the lensing magnification corresponds to an intrinsic $\simeq32.5$\,mag at the fiducial magnification.
At \zsn\ this observed wavelength probes rest-frame $\sim3800$\,\AA; for a flat UV slope the corresponding absolute magnitude limit is $M_{\rm UV}\gtrsim-13$. 
The transient itself is an unresolved point source, with no excess flux over the empirical-PSF expectation in larger apertures, leaving no clear sign of an extended host beneath the SN.
ALMA 2\,mm continuum imaging of the field (M.~VanWyngarden et al. in prep.) also shows no emission at the SN position (Figure~\ref{fig:host}, top right), with a $2\sigma$ limit of $71\,\mu$Jy\,beam$^{-1}$ ($6.3\,\mu$Jy\,beam$^{-1}$ de-lensed). Correcting for the magnification and assuming a $T_{\rm d}=35$\,K modified blackbody, this implies $L_{\rm IR}\lesssim8\times10^{10}\,L_{\odot}$ and an obscured star-formation rate (SFR) of ${\rm SFR}_{\rm IR}\lesssim14\,M_{\odot}\,{\rm yr}^{-1}$, so a fully obscured dusty starburst host is unlikely.

Figure~\ref{fig:host} places this limit in the $M_{\rm UV}$--redshift plane, where it falls well below the reach of unlensed imaging at any depth now practical, in the dwarf-galaxy regime accessible only through strong lensing.
The limit is fainter than the host of SN Eos ($M_{\rm UV}=-14.4\pm0.3$ at $z=5.13$), the faintest and most distant host among spectroscopically confirmed CCSNe \citep{asada2026eos}, so \snh marks a second lensed system whose progenitor environment sits below the general blank-field detection limit.
Convolving the UV luminosity function with the SFR and the magnification-dependent effective volume, \citet{asada2026eos} find that the host of a high-redshift CCSN should most often be a galaxy near $M_{\rm UV}\simeq-19$, and that a host as faint as the one they detect stays unlikely unless the core-collapse rate per unit star formation is enhanced in the metal-poor environments typical of faint galaxies.
Our non-detection of the host is consistent with that picture at a similar high redshift. 
The host is also undetected in emission.
No narrow \ha\ component appears on top of the broad SN line. Injection--recovery tests in the same manner as \citet{asada2026eos}, fitting a flexible pseudo-continuum together with an unresolved line at the systemic redshift, bound a host \ha\ line to $<4.8\times10^{-19}$\,erg\,s$^{-1}$\,cm$^{-2}$ and a host \oiii\,$\lambda5007$ line to $<1.2\times10^{-18}$\,erg\,s$^{-1}$\,cm$^{-2}$ (both $2\sigma$). We adopt the limit on the SFR based on the rest-UV non-detection, ${\rm SFR}\lesssim0.012\,M_{\odot}\,{\rm yr}^{-1}$ for $\mu=11.2$ \citep[conversion from][]{kennicutt1998}, while the \ha\ limit independently gives a consistent ${\rm SFR}({\rm H}\alpha)\lesssim0.03\,M_{\odot}\,{\rm yr}^{-1}$.
The 2D spectrum likewise shows no spatially offset \ha\ emission within $\pm1\arcsec$ along the slit.
Without the cluster magnification \snh would have been recorded as one of the hostless transients now routinely found in deep \jwst\ imaging \citep[e.g.,][]{decoursey2025jades,toshikage2026}, a population usually attributed to hosts below the detection limit.
A future host detection, through deeper imaging at this position, at the counter-image positions where the magnification can be higher, or through Ly$\alpha$ in the deep optical IFU data (e.g., VLT/MUSE, {\it Keck}/KCWI), would simultaneously constrain the progenitor environment and hand the lens models the multiply imaged host they currently lack.

\subsection{A $z\approx16$ impostor} \label{sec:impostor}

The discovery photometry of \snh initially supported a very different interpretation.
With a $\gtrsim2$\,mag break across F150W--F200W (Table~\ref{tab:phot}), detections in all eight redder bands, and no host, standard galaxy-template fitting of the discovery SED with \texttt{EAZY} \citep{brammer2008} returned a clean photometric solution at $z\approx16$, which we took as a promising candidate until the transient nature of the source was recognized.
Three observations dismantled the high-redshift interpretation within a day: faint but positive flux in the dropout band F150W, the flat F200W$-$F210M color that Lyman-break templates struggle to reproduce, and, decisively, variability with respect to the 2024 pre-explosion imaging. Lensing geometry independently favored $z\simeq3$--4.
The same evidence excludes the static red point sources that contaminate dropout searches, Galactic brown dwarfs and the compact ``little red dots'', which show neither month-timescale variability at this level nor broad hydrogen P-Cygni features.

\begin{figure*}
\centering
\includegraphics[width=0.95\linewidth]{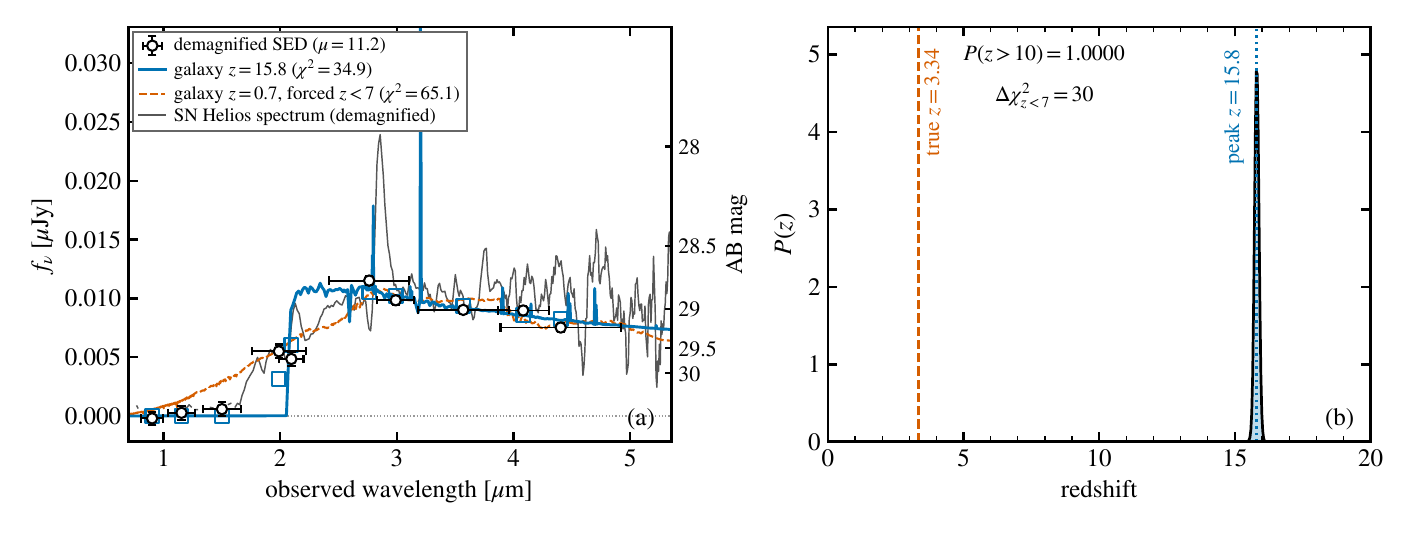}
\caption{
The $z\approx16$ impostor experiment.
Left: the discovery-epoch SED of \snh demagnified by $\mu=11.2$ and evaluated at an assumed survey depth of $30.25$/$30.75$\,mag ($5\sigma$, SW/LW); the fluxes are the exact demagnified measurements, with no synthetic noise added.
Horizontal bars show the effective widths of the NIRCam filters; the right-hand axis marks the corresponding AB magnitudes (the nonuniform spacing follows from the linear flux axis).
The best free-redshift galaxy template (blue, $z=15.8$) and the best solution forced to $z<7$ (orange dashed, $z=0.7$) are shown, together with the demagnified NIRSpec spectrum of \snh (gray), the actual $z=3.34$ SN that these photometric points represent.
Right: the free-redshift photometric posterior of the demagnified SED, which is single-peaked at $z\simeq16$ with $P(z>10)>0.9999$; the best solution forced to $z<7$ is disfavored at $\Delta\chi^{2}=30.1$.
Galaxy templates only, flat redshift prior; see Section~\ref{sec:impostor} for the disclosed assumptions.
}
\label{fig:impostor}
\end{figure*}

Figure~\ref{fig:impostor} demonstrates how convincing the impostor would have been without the lens.
We divide the observed discovery-epoch fluxes by the fiducial $\mu=11.2$, which places the source at $28.7$\,mag in F277W with a clean, fully undetected F150W dropout, and evaluate the SED at an assumed survey depth of $30.75$\,mag ($5\sigma$ point source), i.e.\ the F277W detection has ${\rm S/N}\simeq32$. 
Fitting this SED with \texttt{EAZY}, using the standard \texttt{sfhz} galaxy templates and the redshift free over $0.05<z<20$, returns a sharp, single-peaked solution at $z=15.8$ ($\chi^{2}=34.9$ over ten bands; $P(z>10)>0.9999$ under a flat prior), while the best solution forced to $z<7$ (a red galaxy at $z=0.7$) is disfavored at $\Delta\chi^{2}=30.1$; the true SN spectrum, overplotted in Figure~\ref{fig:impostor}, shows how a Balmer break plus \ha emission at $z=3.34$ conspires to imitate a Lyman break at $z\approx16$.
A quarter magnitude shallower gives $\Delta\chi^{2}\simeq19$, and another half magnitude brings it down to $\simeq8$, so deep surveys reaching $30$--$31$\,mag are precisely in the regime where this impostor is both well-detected and confidently misclassified.
Deep blank-field surveys have already reported photometric high-redshift candidates that later proved to be transients \citep{decoursey2025jades}, and the demagnified \snh sits at $28.7$\,mag in F277W, well within their reach.
The lensing cluster played a role here in making the transient bright enough to be caught, monitored, and spectroscopically unmasked, which turns \snh into a confirmed, empirical template of this contaminant class that the community can fit against future dropout candidates.
Lensing clusters subtend a tiny solid angle compared to the blank-field survey area, so finding one such impostor behind a single cluster suggests that wide unlensed searches face a proportionally larger contamination.
Indeed, a similar but intrinsically more luminous case has just been reported in a blank field, a $z\sim14$ candidate reinterpreted as a SN Ia candidate at $z\sim4.3$ \citep{toshikage2026}.
Medium-band coverage (through the flat F200W$-$F210M color), photometry at multiple epochs, and difference imaging against any earlier data provide inexpensive discriminants, and the surface density of such impostors ties directly to the volumetric SN II rate; the parent core-collapse rate over $2.8\lesssim z\lesssim5$ has now been measured with \jwst, at $4.1^{+1.5}_{-1.1}\times10^{-4}$\,yr$^{-1}$\,Mpc$^{-3}$ \citep{decoursey2026rates}.

\subsection{Summary}

We have presented the discovery and spectroscopic confirmation of \snh, a strongly lensed Type II SN at $z=3.34$ behind \cluster, identified in VENUS multi-epoch NIRCam imaging and confirmed with a DDT NIRSpec prism spectrum showing broad \ha, \hb, and \hei\ P-Cygni features from a photosphere expanding at $7$--$9\times10^{3}$\,\kms. Comparing to local analogs, we recover a photometrically derived phase estimate for the spectroscopic epoch of $27.5\pm2.8$ days, further corroborated by spectroscopic comparisons of the \ha feature to the CSP sample of SNe II.
Five independent lens models predict another image of the SN appearing $2$--$6$\,yr from now, making \snh the highest-redshift cosmologically-useful SN known and a realistic anchor for time-delay cosmography at $z>3$.
The host galaxy is undetected down to a de-lensed $M_{\rm UV}\gtrsim-13$, fainter than any known CCSN host.
And had it not been magnified, monitored, and spectroscopically unmasked, the discovery-epoch SED of \snh would stand today as a convincing $z\approx16$ candidate; its spectrum now serves as an empirical template for vetting the dropout candidates of high-redshift galaxy searches, and we make it available with the \texttt{EAZY} template sets \citep{brammer2008}\footnote{\url{https://github.com/gbrammer/eazy-photoz}}.

\vspace{5mm}
\noindent {\bf Acknowledgments.}
SF acknowledges support from the Dunlap Institute, funded through an endowment established by the David Dunlap family and the University of Toronto, and from an NSERC discovery grant (RGPIN-2026-07931; DGECR-2026-00232). 
RAW acknowledges support from NASA JWST Interdisciplinary Scientist grants NAG5-12460, NNX14AN10G and 80NSSC18K0200 from GSFC.
PD warmly acknowledges support from an NSERC discovery grant (RGPIN-2025-06182).
KK acknowledges the support by JSPS KAKENHI Grant Numbers JP22H04939, JP23K20035, and JP24H00004.
We acknowledge the support of the Canadian Space Agency (CSA) [25JWGO4A18]. 

\paragraph*{Use of generative AI:}
We acknowledge the use of a generative AI tool to refine the text and the code used in this work. 
All figures and quoted values were reproduced and verified by the authors.

\bibliographystyle{apj}
\bibliography{apj-jour,reference,conor.bib}

\appendix

\section{CSP Spectroscopic Comparison}

\begin{figure*}
\centering
\includegraphics[width=0.4\linewidth]{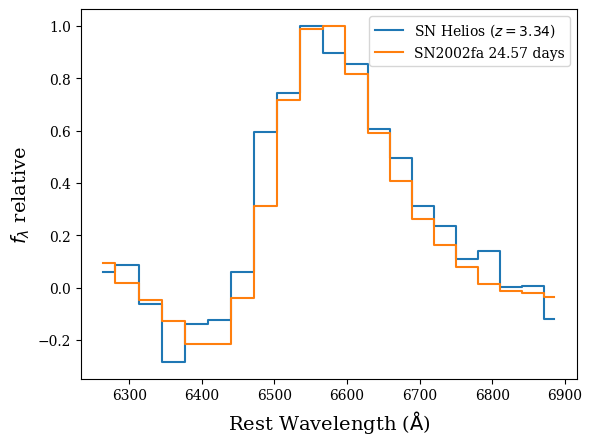}
\includegraphics[width=0.5\linewidth]{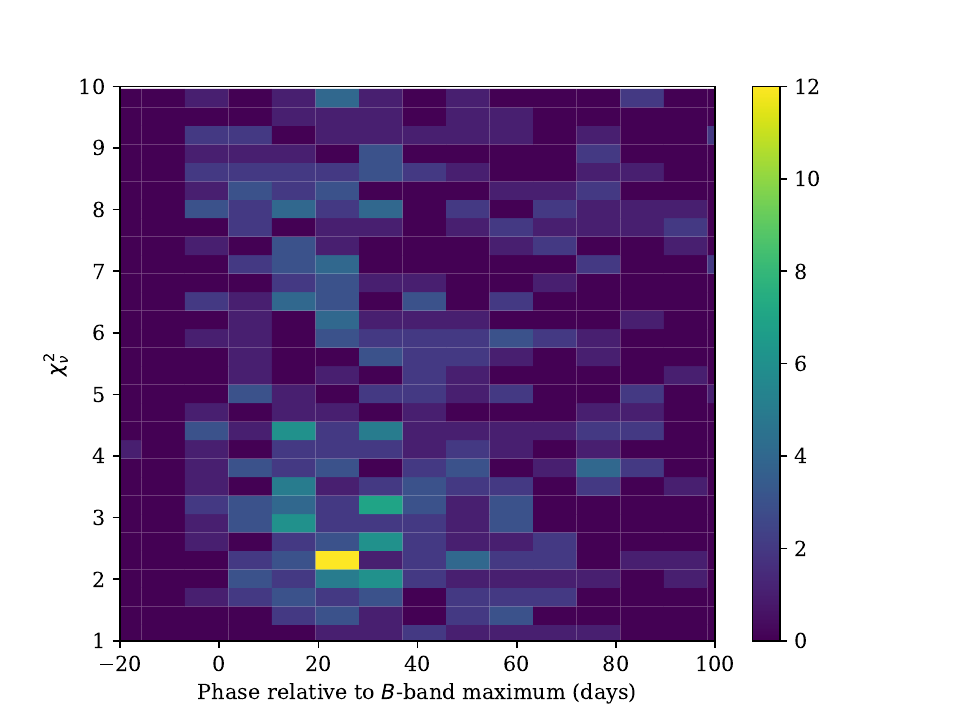}

\caption{\textit{\textbf{Left:}} Example comparison between the SN~2002fa +24.57-day \ha feature from \citet{Gutierrez_2017} with \snh. This specific comparison showed a good match, with a $\chi^2_{\nu}$ value of $\sim2.3$. \textit{\textbf{Right:}} A 2-D histogram of the $\chi^2_{\nu}$ values from the CSP sample, for $\chi^2_{\nu}<10$ (representing 447 spectra across 108 SNe II). By far, the highest density of low-$\chi^2_{\nu}$ value comparisons falls within the $20-30$ day post-peak phase range.
}
\label{fig:csp_spec} 
\end{figure*}

In Figure~\ref{fig:csp_spec}, we show an example of the \ha feature analysis that we use to estimate the spectroscopic phase of \snh. We can see a close match in the comparison of this feature between \snh and SN 2002fa at +24.57 days, with both the velocity shift and FWHM agreeing quite well for both the emission and absorption components of the \ha profile. For the total sample, we find that the largest overdensity of quality fits ($\chi^2_{\nu}\approx2$), falls within the $20-30$ day phase range. Although there is clearly large scatter in terms of phase, this result agrees strongly with our light curve-fitting results, which produce an estimated spectral phase of $27.5\pm2.8$ days relative to \textit{B}-band peak. Further spectroscopic observations during the reappearance of \snh would allow us to better estimate the phase of this classification spectrum while further opening the door to the expanding photosphere method and spectroscopic time delays.

\section{SN 1999em and SN 2005cs light curve comparisons}

\begin{figure*}
\centering
\includegraphics[width=0.45\linewidth]{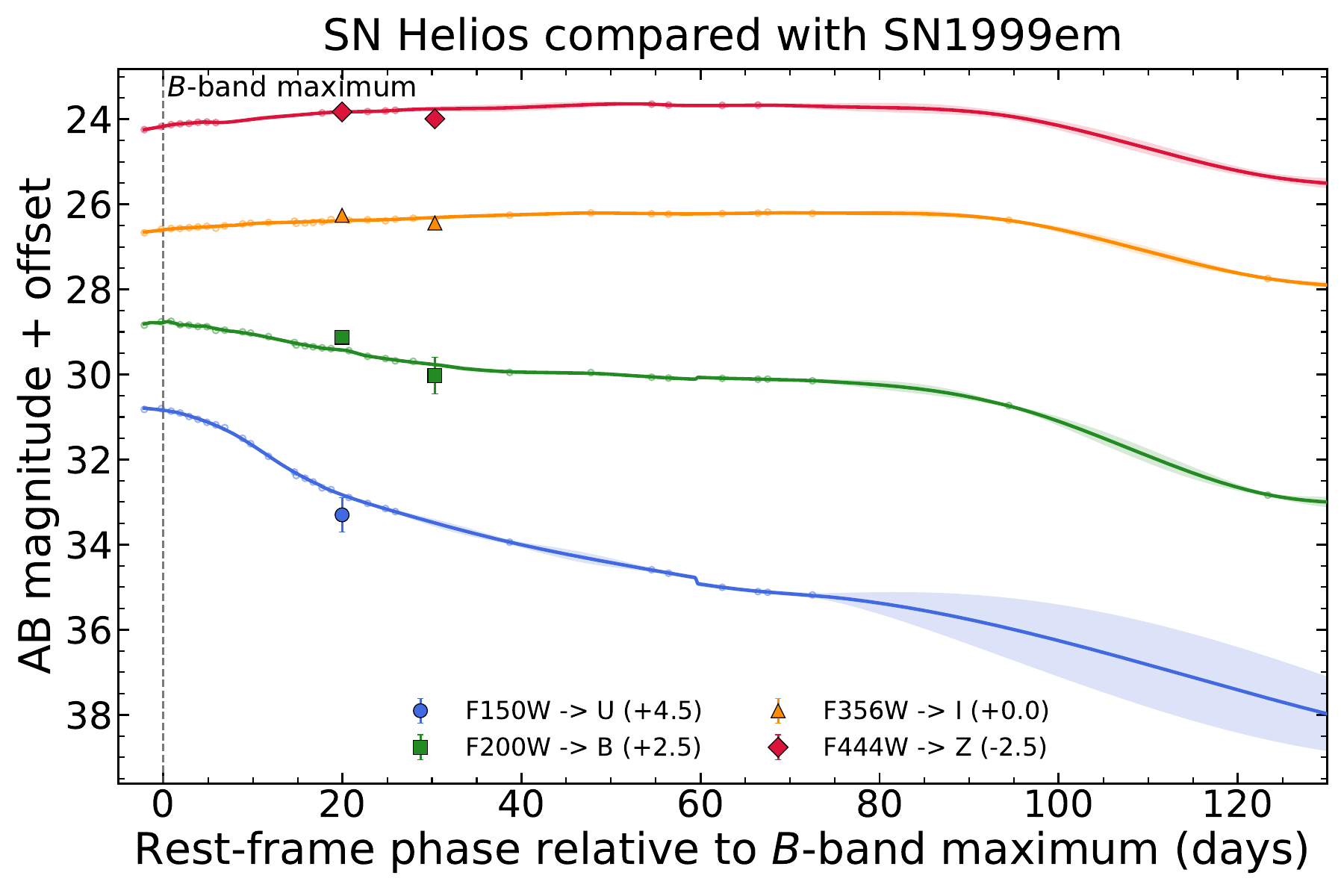}
\includegraphics[width=0.45\linewidth]{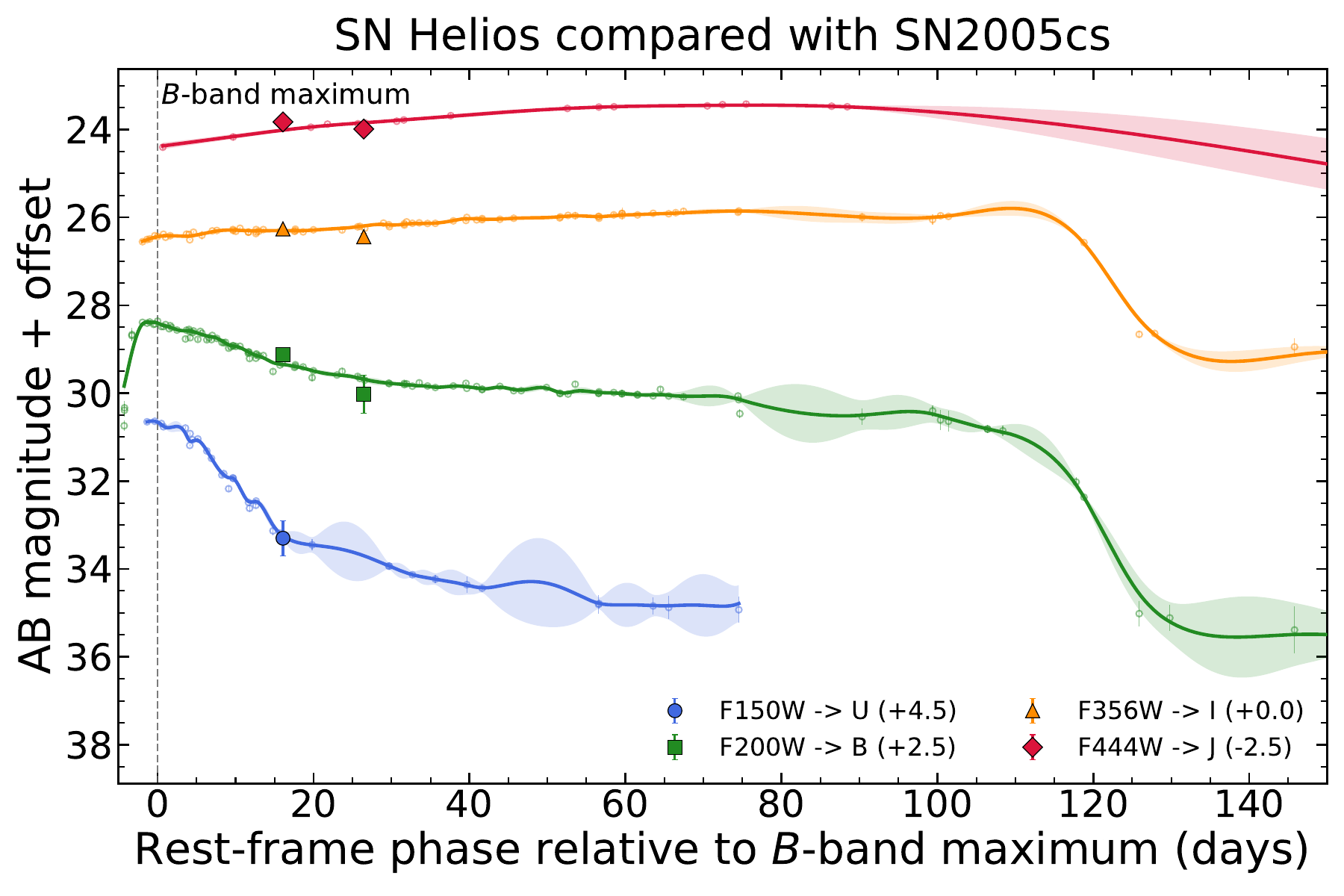}

\caption{Light curve matches for SN 1999em and SN 2005cs, the analysis methodology behind which is described in Section~\ref{sec:lightcurve}, and the assumptions in the fits described in Table~\ref{tab:template_assumptions}. The measured phase of the resulting fits are $30.3\pm5.2$ and $26.4\pm4.6$ days relative to the \textit{B}-band maximum for SN~1999em and SN~2005cs, respectively. Despite the clearly shallower evolution in their light curves, these results are still consistent with SN~2013ej within $1\sigma$ and with the spectroscopic estimate.
}
\label{fig:other_lcs} 
\end{figure*}

In Section~\ref{sec:lightcurve}, we discuss the method by which we estimate the phase of the DDT observations using matches to low-redshift SN II analogs. In Figure~\ref{fig:lightcurve}, we show the best-matching light curve, which is to the fast-declining SN~2013ej. Figure~\ref{fig:other_lcs} shows the light curve fits to the two other low-z comparison objects, SNe 1999em and 2005cs. Both exhibit a much longer plateau than SN~2013ej, with F356W and F444W curves that actually increase in brightness between the two epochs of \snh, rather than decrease as the \snh light curve does. Our upcoming epoch of photometry in November-December should result in a datapoint on the optical plateau, which should greatly help the final estimate of the time delay with respect to the reappearing image. Despite the apparent difference in decline rates between these SNe and the light curve of \snh, the phases at which these observations were taken are such that the best-fit phase is still post-optical peak and pre-optical plateau. If the observations showed less evolution, they would be favored by the plateau, while a faster decline would certainly relate to the fall of the plateau. In this way, the light curve of \snh was sampled at an optimal time for this phase (and eventually time delay) estimation.

\section{An empirical redshift cross-check with SN Eos}
\label{app:eos}

\begin{figure}
\centering
\includegraphics[width=\linewidth]{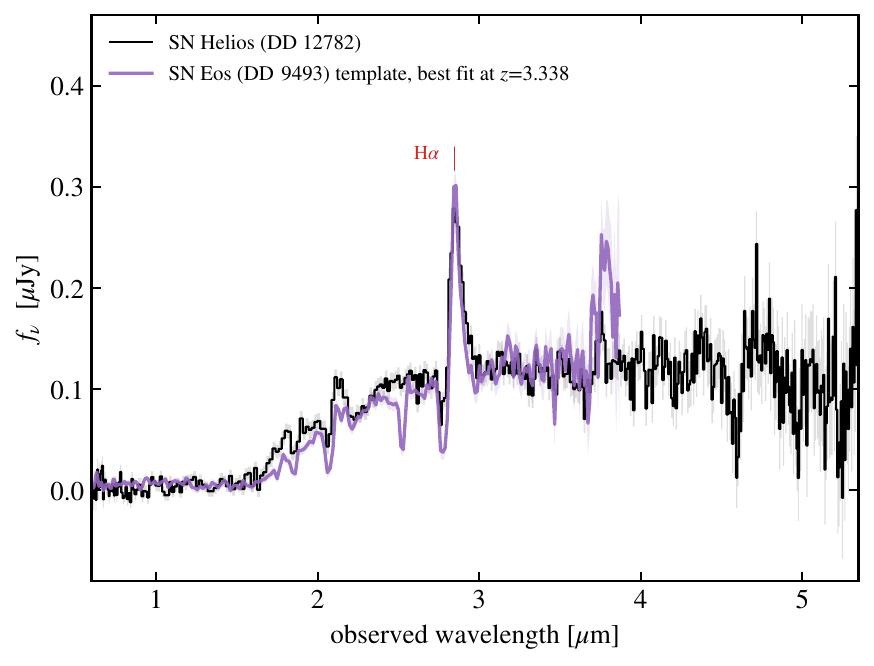}
\caption{
Prism spectrum of \snh\ (black) with the prism spectrum of SN Eos (purple) shifted in $(1+z)$ and rescaled to it, with redshift and amplitude free. The best match is at $z=3.338$. The shaded band shows the propagated SN Eos uncertainty, and the vertical line marks \ha\ at that redshift. SN Eos has no coverage beyond $3.85$\,$\mu$m in this frame.
}
\label{fig:eos}
\end{figure}

SN Eos at $z=5.13$ \citep{coulter2026eos,asada2026eos} is the only other CCSN with a high signal-to-noise NIRSpec prism spectrum at a comparable redshift, taken in the same S200A1 prism mode, so it carries the same line-spread function and the same calibration path. Figure~\ref{fig:eos} compares it with \snh\ after shifting it in $(1+z)$ and rescaling, with redshift and a single amplitude the only free parameters. The best match falls at $z=3.338$, close to the value adopted from the line measurements.

Note that this agreement should be read as a consistency check rather than as an independent measurement. Neither spectrum shows a significant narrow host line at \ha, so the comparison matches two broad P-Cygni profiles against each other, and the emission peak of such a profile does not sit at the systemic wavelength. Measuring both peaks against their own systemic redshifts gives $+936\pm343$\,\kms\ for \snh\ and $+376\pm118$\,\kms\ for SN Eos, whose systemic redshift comes from a narrow host \ha\ line. The two offsets differ by $560\pm363$\,\kms, which propagates to a systematic of $\delta z\simeq0.008$ on the redshift recovered this way.

\end{document}